\documentclass[%
 reprint,
nofootinbib,
 amsmath,amssymb,
 aps,
prd,
showkeys,
]{revtex4-2}

\usepackage{verbatim}
\usepackage{float}

\usepackage{graphicx}
\usepackage{dcolumn}
\usepackage{bm}
\usepackage{hyperref}
\hypersetup{
    colorlinks=true,
    linkcolor=blue,
    filecolor=magenta,      
    urlcolor=cyan
}
\usepackage{color,soul}

\usepackage{soul}
\usepackage{color}
\usepackage{ulem}
\usepackage{slashed}
\usepackage{appendix}
\usepackage{multirow}

\newcommand{\be}{\begin{equation}}
\newcommand{\ee}{\end{equation}}

\begin{document}


\title{Confronting \mbox{\boldmath $K^-d$} dynamics with kaonic atom measurements \\ within a novel unitary framework}

\author{Albert Feijoo}
 \email{edfeijoo@ific.uv.es}
 
\affiliation{Departamento de F\'\i sica Te\'orica and IFIC, Centro Mixto Universidad de Valencia-CSIC, Institutos de Investigaci\'on de Paterna, Aptdo. 22085, E-46071 Valencia, Spain}

\author{\`Angels Ramos}
 \email{ramos@fqa.ub.edu}
 
\affiliation{Departament de Fı\'isica Qu\`antica i Astrof\'isica and Institut de Ci\`encies del Cosmos (ICCUB), Facultat de F\'isica, Universitat de Barcelona, Mart\'i i Franqu\`es 1, 08028 Barcelona, Spain}

\author{Juan Torres-Rincon}
 \email{torres@fqa.ub.edu}
 
\affiliation{Departament de F\'isica Qu\`antica i Astrof\'isica and Institut de Ci\`encies del Cosmos (ICCUB), Facultat de F\'isica, Universitat de Barcelona, Mart\'i i Franqu\`es 1, 08028 Barcelona, Spain}

\date{\today}

\begin{abstract}

We study the low-energy interaction of negative kaons with deuterons using chiral $\bar{K}N$ interaction models (OR and BCN) as input. The three-body $K^-d$ amplitude is constructed within the Fixed-Center Approximation to the Faddeev equations, where the unitary constraints missing in the conventional approach are formally implemented with the incorporation of coherent multiple-scattering processes. The unitary procedure is extended to coupled channels including charge-exchange processes, as well as isospin-breaking effects through the use of physical particle masses. This provides a consistent description of the low-energy $K^-d$ amplitude and allows for the determination of the scattering length and the effective range. The resulting framework is then used to assess the compatibility of the BCN $\bar{K}N$ interaction model with the recent SIDDHARTA-2 measurements of kaonic hydrogen and kaonic deuterium through a bootstrap analysis. The procedure leads to a readjustment of the BCN parameterization which produces visible effects in several isospin-sensitive $\bar{K}N$ scattering observables, hence highlighting the additional information provided by the new kaonic-atom measurements to better constrain the poorly known isovector component of the ${\bar K}N$ interaction.
\end{abstract}

\keywords{kaon-nucleon interaction, kaonic deuterium, Faddeev equations, unitarity.}
\maketitle


\section{Introduction}
\label{sec:intro}

The $K^-d$ system has been actively investigated for over half a century as a way to test the properties of the kaon--deuteron interaction and its connection with the underlying elementary $K^- p$ and $K^- n$ ones (${\bar K} N$ in short). The studies originated with early Faddeev calculations of low-energy $K^-d$ cross sections~\cite{Hetherington:1965zza}. Subsequent refinements explicitly incorporated couplings to hyperonic $\pi NY$ decay channels, although these were shown to exert a rather negligible influence on the final observables~\cite{Toker:1981zh,Torres:1986mr}. 

Since obtaining $K^-d$ cross section data at low energies is technically challenging,  the properties of the $K^-d$ interaction, such as the scattering length, must be extracted from kaonic deuterium measurements, a goal that has been successfully achieved recently by the SIDDHARTA-2 collaboration~\cite{Bazzi:2026gri}. Simultaneously the ALICE collaboration has provided new insights into the scattering lengths of the $K^{\pm}d$ system through the analysis of measured correlation functions in femtoscopy experiments~\cite{ALICE:2026pxr}. 

On the theoretical side, right after the appearance of unitary chiral effective field theories~\cite{Kaiser:1995eg,Oset:1997it,Jido:2002zk}, which accurately describe $\bar{K}N$ scattering and dynamically generate the $\Lambda(1405)$ resonance, calculations of the $K^-d$ scattering length were performed~\cite{Kamalov:2000iy,Deloff:1999gc,Bahaoui:2003xb,Meissner:2006gx} either solving the Faddeev equations or within their simpler Fixed-Center Approximation (FCA) version, which neglects nucleon recoil effects.  Detailed comparisons with full Faddeev calculations have shown that the conventional FCA provides a reasonable approximation to the $K^-d$ scattering length, with deviations typically at the 10--25\% level~\cite{Gal:2006cw}, depending on the treatment of the underlying three-body dynamics \cite{Deloff:1999gc,Bahaoui:2003xb,Gal:2006cw}. Recoil corrections were also found to be reasonably small \cite{Baru:2009tx,Mai:2014uma}. A few years after the development of the chiral models, the more precise measurement of kaonic hydrogen by the SIDDHARTA collaboration~\cite{SIDDHARTA:2011dsy} placed tighter constraints on the $K^-p$ interaction and prompted renewed interest in the study of the $K^-d$ system~\cite{Doring:2011xc,Mizutani:2012gy,Shevchenko:2011ce,Revai:2012fx,Shevchenko:2014uva,Revai:2016muw,Liu:2020foc,Duerinck:2025pbj}.

Knowledge of the $K^- d$ scattering length allows for a straightforward evaluation of the energy shift and width of the 1s level of kaonic deuterium employing the original Deser formula~\cite{Deser:1954vq}, the corrected Deser-Trueman one~\cite{Meissner:2004jr,Meissner:2005bz}, or the summed-up expression~\cite{Baru:2009tx,Shevchenko:2021swf}. The 1s state of kaonic deuterium has also been directly calculated, either from two-body reduced or proper three-body calculations~\cite{Barrett:1999cw,Hoshino:2017mty,Revai:2016muw,Duerinck:2025pbj}, showing deviations of a few-percent with the summed-up estimates.

Building on this established framework, our previous work~\cite{Ramos:2025ibe} successfully employed the elementary ${\bar K}N$ amplitudes of the so-called OR model~\cite{Oset:1997it,Jido:2002zk} within the FCA to obtain, not only the $K^- d$ scattering length, but also the $K^- d$ femtoscopy correlation functions, which compared favorably with experiment~\cite{ALICE:2026pxr}. However the recent precise measurement of the 1s level in kaonic deuterium by the SIDDHARTA-2 collaboration~\cite{Bazzi:2026gri} motivates a reassessment of the framework used to derive the $K^-d$ scattering amplitude from the elementary $\bar{K}N$ interactions.

The present study improves upon the previous effort~\cite{Ramos:2025ibe} in two significant aspects. First, we employ a state-of-the-art next-to-leading order (NLO) chiral interaction model, namely the BCN model~\cite{Feijoo:2018den}, which was precisely fitted to the 1s level of kaonic hydrogen alongside other key isospin observables. Second, we formally incorporate the unitary constraints that are inherently absent from the traditional FCA framework by accounting for the coherent propagation of the kaon with the deuteron cluster~\cite{Encarnacion:2026zas,Ikeno:2025bsx,Agatao:2025ckp}. We extend this unitarization scheme to comprehensively account for the effects of coupled channels, explicitly including exchange processes and isospin-breaking effects arising from the use of physical particle masses. These effects are not included in the recent $\bar K^0d$ femtoscopic study~\cite{Bayar:2026ted} and, although they are found to have a relatively limited impact in the present work, their formal incorporation leads to a more general framework applicable to systems governed by coupled-channel dynamics. The resulting $K^-d$ amplitude satisfies strict unitarity, thus admitting a well-defined effective range expansion. This, in turn, allows us to determine not only the scattering length but also the effective range, providing essential parameters for direct comparison with future femtoscopic measurements.

We conclude this work by investigating the constraining power of the recent SIDDHARTA-2 measurements of the 1s level of kaonic deuterium~\cite{Bazzi:2026gri} and the improved 1s kaonic-hydrogen measurement~\cite{Bazzi:2026zmc} on the BCN model. To this end, we perform a bootstrap analysis in which the BCN parameters are allowed to readjust around their original fitted values, with their individual uncertainties retained as priors. The simultaneous consideration of kaonic hydrogen and kaonic deuterium provides complementary sensitivity to the isospin structure of the $\bar{K}N$ interaction, particularly to its poorly constrained isovector component. We consider two prescriptions for obtaining the kaonic-atom 1s energy shift and width from the $K^- d$ amplitude, corresponding to treatments in which electromagnetic corrections are included either perturbatively, at first order, or through the fully resummed expression. The bootstrap results are found to be broadly consistent with the SIDDHARTA-2 measurements, although some differences emerge in the kaonic-deuterium observables depending on the treatment of electromagnetic corrections. The resulting parameter variations remain moderate, but they are sufficient to induce discernible changes in several isospin-sensitive scattering observables.  
These findings illustrate the potential of the new kaonic-atom measurements to provide additional constraints on the low-energy $\bar{K}N$ interaction.

The paper is organized as follows. In Sect.~\ref{sec:ampl}, the derivation of the $K^-d$ interaction from the elementary ${\bar K}N$ amplitudes is described. First, the essential points of the FCA are recalled, followed by the implementation of elastic unitarity constraints including charge-exchange processes. Results for $K^-d$ low-energy observables are presented in Sect.~\ref{sec:results}, together with the bootstrap analysis aimed at establishing the compatibility of the BCN model with the new SIDDHARTA-2 measurements of kaonic hydrogen and deuterium. Finally, in Sec.~\ref{sec:conclusions} the work is summarized and key findings are highlighted.

\section{Kaon-deuteron scattering amplitude}
\label{sec:ampl}

In this section, we describe how the $K^- d$ scattering amplitude is obtained in terms  of the elementary ${\bar K}N$ amplitudes within the so-called FCA to the Faddeev equations. We subsequently introduce additional extensions to enforce exact unitary constraints in the final scattering amplitude.

\subsection{Fixed-Center Approximation}

By ignoring the target nucleon recoil within the deuteron during the kaon interactions, the FCA offers an elegant route to express the $K^- d$ scattering length, $A_{K^- d}$, directly in terms of threshold $\bar{K}N$ amplitudes. In turn, employing $A_{K^- d}$ in either the corrected~\cite{Meissner:2005bz} or the summed-up Deser-Trueman formula~\cite{Baru:2009tx,Shevchenko:2021swf} subsequently connects these elementary scattering parameters to the strong-interaction energy shift and width of the 1s level of kaonic deuterium.

As explicitly shown in our previous work~\cite{Ramos:2025ibe}, the FCA $K^-d$ amplitude, $T^\mathrm{FCA}_{K^- d}$, is obtained from 
    \begin{equation}
        T_{K^-d}^\mathrm{FCA} = T_p + T_n \ ,
    \end{equation}
where the scattering amplitudes $T_p$ and $T_n$ correspond to the Faddeev partitions that contain the processes in which the $K^-$ collides first on a proton and a neutron, respectively. These are obtained from the following system of equations,
\begin{equation}
\begin{array}{rcl}
    T_p &=& t_1 + t_1 G_- T_n - t_x G_0 T^\mathrm{x}_n \ , \\
    T_n &=& t_2 + t_2 G_- T_p \ ,  \label{eq:FCAnew} \\
    T^\mathrm{x}_n &=& t_x - t_0 G_0 T^\mathrm{x}_n + 
    t_x G_- T_n  \ , 
\end{array}
\end{equation}
where $T_n^\mathrm{x}$ is the Faddeev partition that describes the $\bar{K}^0nn \to K^-pn$ transition, including multiple scatterings in the intermediate states. The quantities $t_1$, $t_2$, $t_0$ and $t_x$ are the following elementary two-body amplitudes, 
\begin{eqnarray}
   && t_1 \equiv t_{K^-p \to K^-p}, \quad t_2 \equiv t_{K^-n \to K^-n} \ , \nonumber \\
   && t_0\equiv t_{{\bar K}^0 n \to {\bar K}^0 n}, \quad t_x \equiv t_{K^-p \to {\bar K}^0 n} \ . \nonumber
\end{eqnarray}
In this present work, these amplitudes are taken from the lowest-order chiral unitary model developed in Ref.~\cite{Oset:1997it} (OR model), with the improved parameterization of Ref.~\cite{Jido:2002zk}, as well as from the NLO model of Ref.~\cite{Feijoo:2018den} (BCN model), which, apart from the elastic and inelastic ${\bar K}N$ scattering data, is also constrained by the precise energy shift and width of the kaonic hydrogen, measured by the SIDDHARTA Collaboration~\cite{SIDDHARTA:2011dsy}.

In writing Eq.~\eqref{eq:FCAnew}, we have generalized the formalism of Ref.~\cite{Ramos:2025ibe}, by allowing for a distinction between the propagation of $K^-(pn)$ intermediate states ($G_-$) from that of ${\bar K}^0(nn)$ ones ($G_0$)\footnote{The use of this propagator in the kaon double-rescattering contribution, and subsequent rescattering ones, is clearly explained in Appendix A of Ref.~\cite{Malabarba:2024hlv}.} \footnote{This expression also accounts for recoil corrections of the deuteron cluster. The impact of genuine three-body recoil corrections at lowest order in $m_K/m_N$ was analyzed in \cite{Baru:2009tx,Mai:2014uma} and found to be quite limited.},
\begin{equation}
    G_{\rm m} = \int\frac{\mathrm{d}^3 q}{(2\pi)^3}\frac{F_d(\bm{q})}{2\omega_{\rm m}(\bm{q})\left(\sqrt{s_{Kd}}-\omega_{\rm m}(\bm{q})-E_{\rm m}(\bm{q})+{\rm i} \eta \right) }\ ,
        \label{eq:prop}
\end{equation}
where $\sqrt{s_{Kd}}$ is the total energy in the $K^-d$ center-of-mass frame, $E_{\rm m}(\bm{q})=\sqrt{M_{\rm m}^2+\bm{q}^2}$ the two-nucleon cluster energy, and $\omega_{\rm m}(\bm{q})=\sqrt{m_{\rm m}^2+\bm{q}^2}$ the kaon energy, with ${\rm m}=\{-,0\}$ for an intermediate $K^-(pn)$ or ${\bar K}^0(nn)$ state, respectively. The cluster mass $M_{\rm m}$ to be used in this equation reads $M_-=M_d=\tilde{m}_p+\tilde{m}_n$ and $M_-=2\tilde{m}_n$, where the binding energy of the deuteron is absorbed into its nucleon components by reducing their masses as $\tilde{m}_p=\xi m_p$ and $\tilde{m}_n=\xi m_n$, with $\xi=M_d/(m_p+m_n)$~\cite{Montesinos:2024eoy}.

The deuteron form factor, $F_d(\bm{q})$, is given by:
    \begin{equation}
        F_d(\bm{q})=\int  \mathrm{d}^3r \ e^{-{\rm i} \bm{q}\cdot \bm{r}}|\phi_d(\bm{r})|^2 \ ,
    \end{equation}
where $\phi_d(\bm{r})$ is the deuteron wave function. Its $s-$ and $d-$wave components are taken from the analytical expressions given in Ref.~\cite{Mahlein:2023fmx}, with parameters fitted to the realistic deuteron wave function obtained with the Argonnne V18 interaction~\cite{Wiringa:1994wb}. Note that, lacking any information for the structure of a bound two-neutron system $(nn)$, we have assigned to it the same form factor as that for the deuteron $(pn)$. 

After solving the system~\eqref{eq:FCAnew}, the FCA amplitude reads,
    \begin{equation}
  T_{K^-d}^\mathrm{FCA} =   \frac{t_1 + t_2 + 2 t_1 t_2 G_- - \tilde{t}^{\, 2}_x G_0 
 - 2 t_2 \tilde{t}^{\, 2}_x G_- G_0 }{1 - t_1 t_2  G_-^2+ 
 t_2 \tilde{t}^{\, 2}_x G_0 G_-^2 } \ , \label{eq:FCAfinal}
\end{equation}
with $\tilde{t}^{\, 2}_x \equiv t_x^2/(1+t_0G_0)$.

The $K^-d$ amplitude is a function of the invariant $K^-d$ center-of-mass energy $\sqrt{s_{Kd}}$, but it is determined from two-body elementary amplitudes that depend on their own two-body center-of-mass energy,  $\sqrt{s_{K^-p}}$ and $\sqrt{s_{K^-n}}$. Their values are obtained from the energy of the $K^-$ in the frame where the deuteron is at rest, $q^0_- = (s_{Kd}-M_d^2 - m_-^2)/(2M_d)$. 
Then,
\begin{eqnarray}
s_{K^-p}\!\!\!&=&\!\! (q^0_-+\tilde{m}_p)^2- \bm{q}_{-}^2 = m_-^2+ \tilde{m}_p^2+2\tilde{m}_p q^0_- 
\label{eq:two-body-energy1} \ , \\ 
s_{K^-n}\!\!\!&=&\!\! (q^0_-+\tilde{m}_n)^2- \bm{q}_{-}^2 = m_-^2+ \tilde{m}_n^2+2\tilde{m}_n q^0_- \ ,
\label{eq:two-body-energy2}
\end{eqnarray}
with $\bm{q}_{-}^2=(q^0_-)^2-m_-^2$. The variable $s_{K^-p}$ is employed in $t_1$
and its related amplitudes by channel coupling ($t_0$ and $t_x$),  whereas $s_{K^-n}$ is employed in $t_2$.

\subsection{Elastic unitarity}

Recent studies have highlighted the need to incorporate the coherent propagation of the colliding particle and cluster—in our case, the $K^\pm$ and the deuteron—within the FCA framework~\cite{Encarnacion:2025lyf,Ikeno:2025bsx,Agatao:2025ckp}. This refinement ensures that the scattering amplitude satisfies elastic unitarity and establishes consistency with the solution of the corresponding Schrödinger equation for the interacting particle-cluster system.

A procedure to implement this coherent propagation within the FCA was formulated in Ref.~\cite{Ikeno:2025bsx} and rigorously justified in Ref.~\cite{Agatao:2025ckp}, under the assumption of isospin symmetry and omitting exchange contributions. Since our formulation explicitly incorporates exchange processes and accounts for isospin-breaking effects through the use of physical particle masses, the prescription outlined in Refs.~\cite{Ikeno:2025bsx,Agatao:2025ckp} needs to be extended accordingly.

\subsubsection{Reorganizing the diagrams}

\begin{figure*}[htbp]
    \centering      \includegraphics[width=0.80\linewidth]{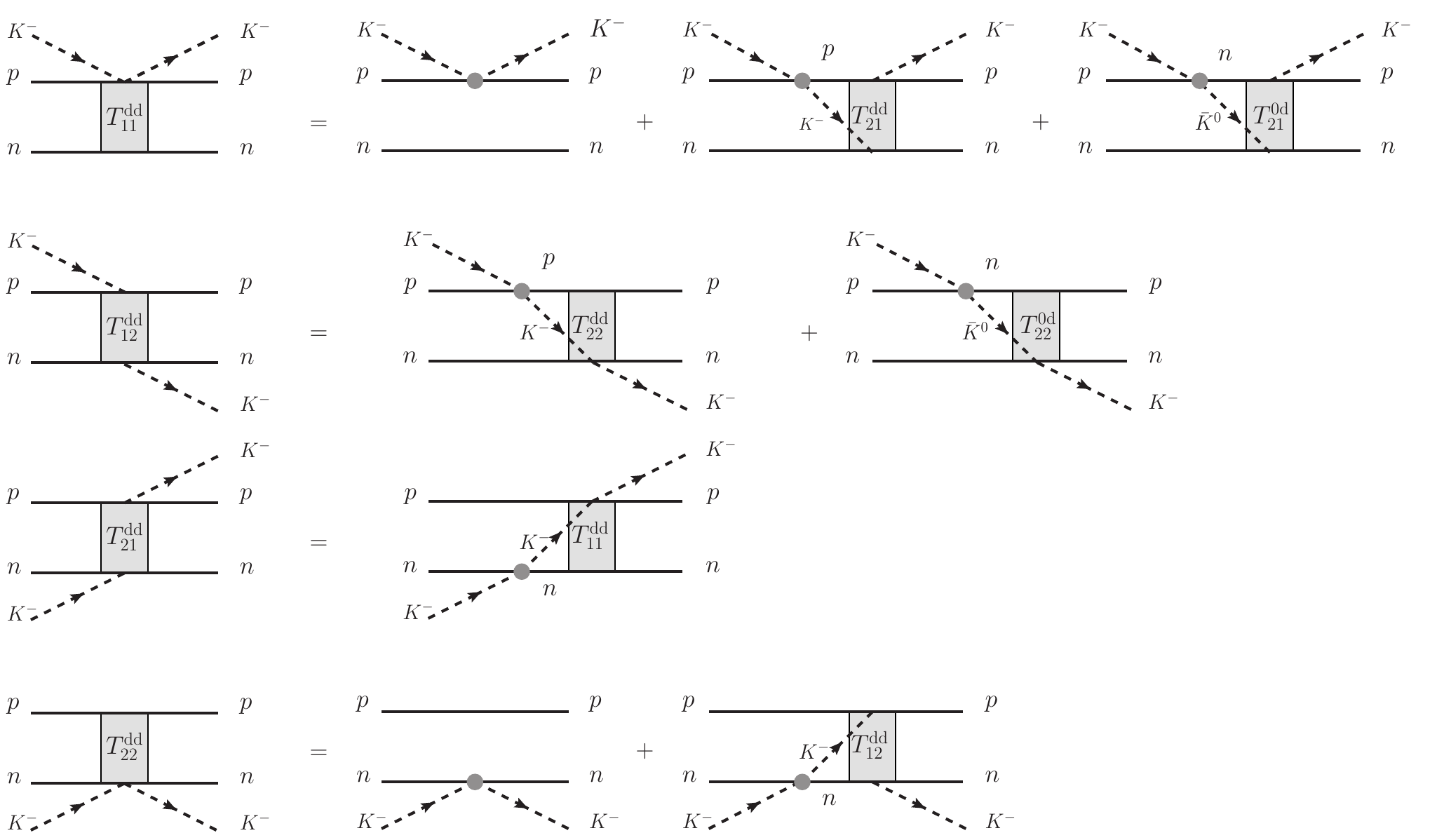}
    \caption{Diagrams representing the transition amplitudes $K^-pn\to K^-pn$, with the $K^-$ interacting first with either particle 1 ($p$) or 2 ($n$) and being emitted from either particle 1 ($p$) or 2 ($n$).}
    \label{fig:dd}
\end{figure*}

\begin{figure*}[htbp]
    \centering      \includegraphics[width=0.7\linewidth]{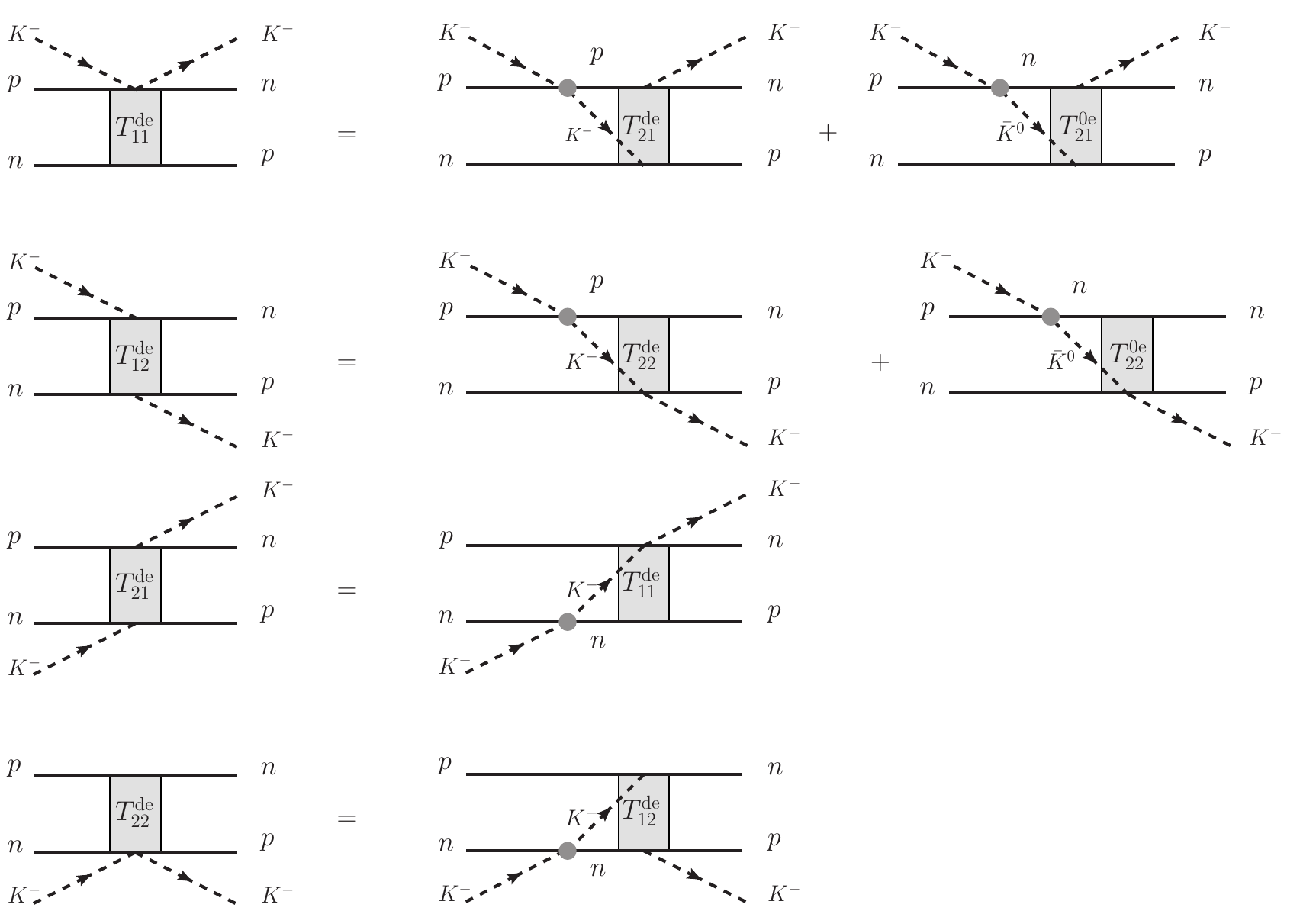}
    \caption{Diagrams representing the transition amplitudes $K^-pn\to K^-np$, with the $K^-$ interacting first with either particle 1 ($p$) or 2 ($n$) and being emitted from either particle 1 ($n$) or 2 ($p$).}
    \label{fig:de}
\end{figure*}

\begin{figure*}[htbp]
    \centering      \includegraphics[width=0.8\linewidth]{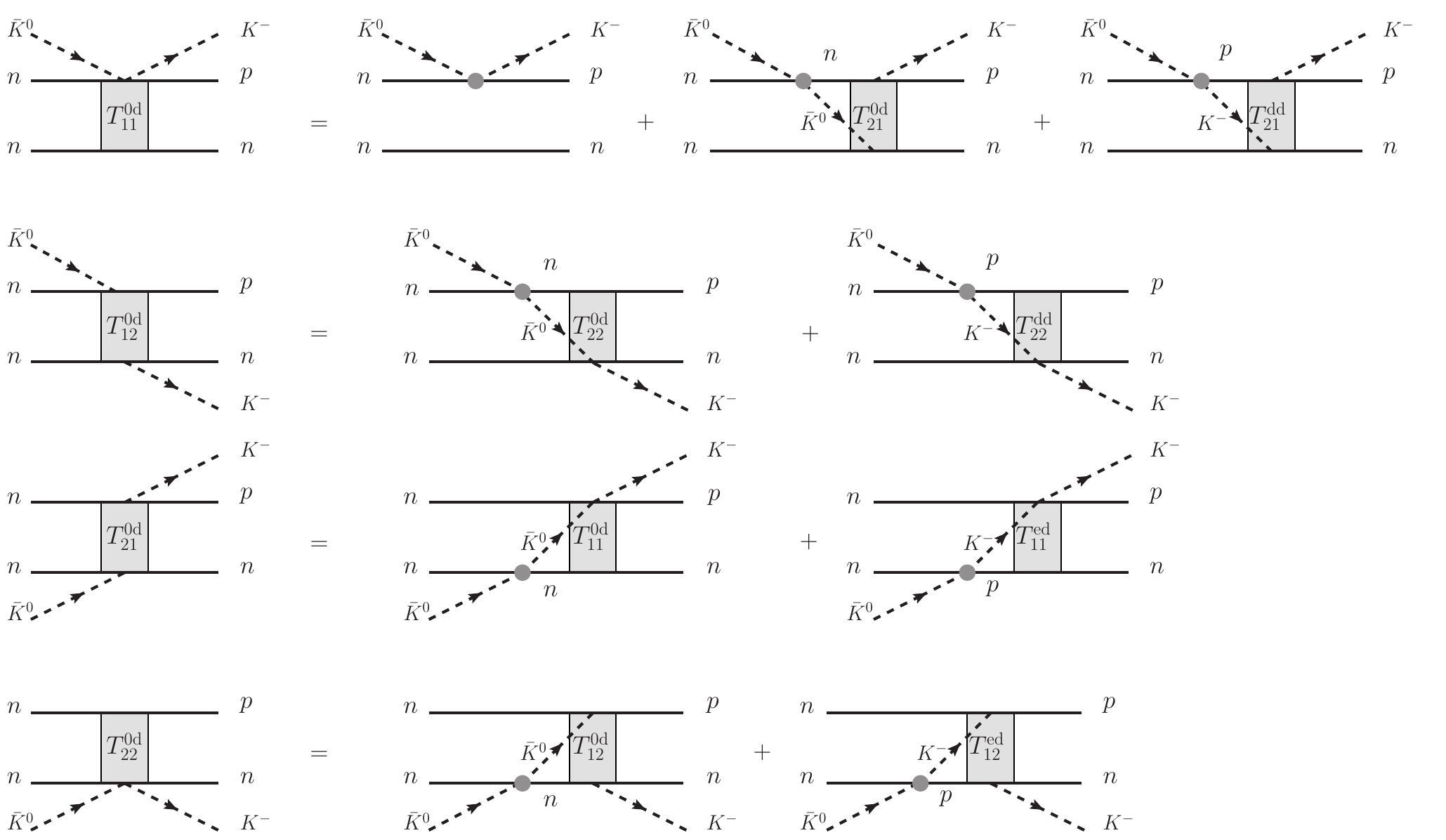}
    \caption{Diagrams representing the transition amplitudes ${\bar K}^0 nn\to K^-pn$, with the ${\bar K}^0$ interacting first with either particle 1 ($n$) or 2 ($n$) and the $K^-$ being emitted from either particle 1 ($p$) or 2 ($n$).}
    \label{fig:0d}
\end{figure*}

\begin{figure*}[htbp]
    \centering      \includegraphics[width=0.8\linewidth]{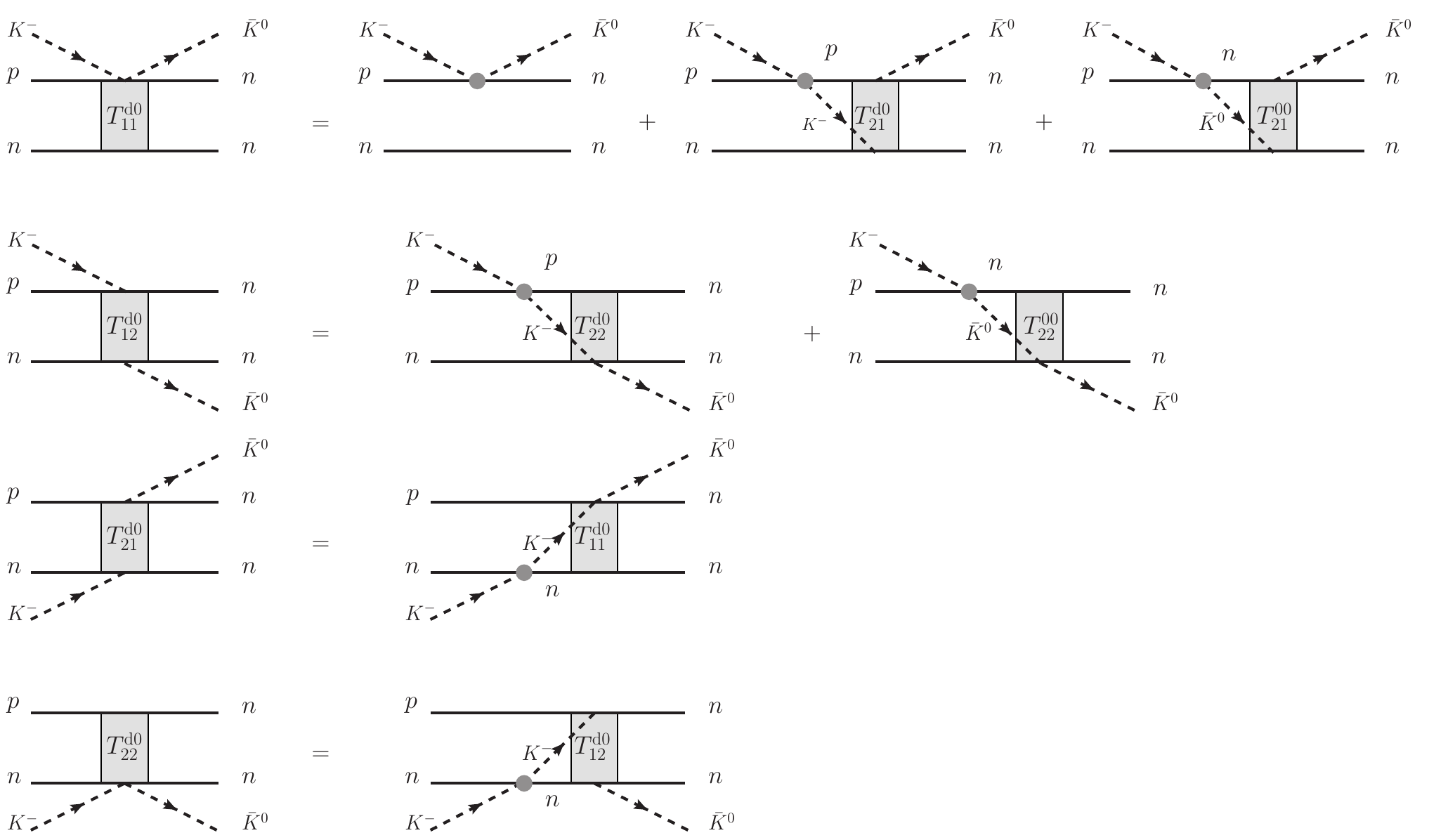}
    \caption{Diagrams representing the transition amplitudes $K^-pn \to {\bar K}^0 nn$, with the $K^-$ interacting first with either particle 1 ($p$) or 2 ($n$) and the ${\bar K}^0$ being emitted from either particle 1 ($n$) or 2 ($n$).}
    \label{fig:d0}
\end{figure*}

\begin{figure*}[htbp]
    \centering      \includegraphics[width=0.8\linewidth]{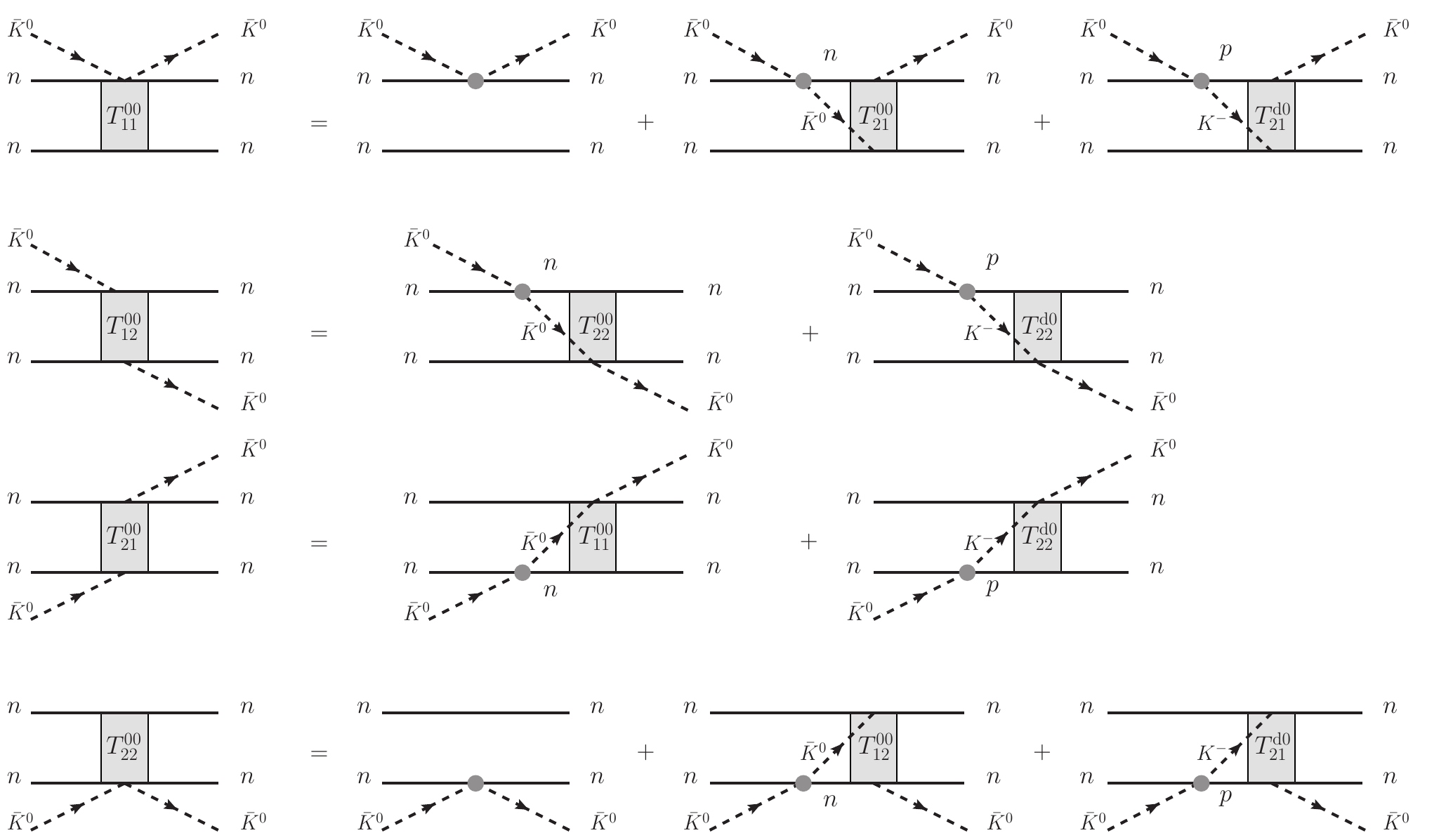}
    \caption{Diagrams representing the transition amplitudes ${\bar K}^0 nn \to {\bar K}^0 nn$, with the ${\bar K}^0$ interacting first with either particle 1 ($n$) or 2 ($n$) and the ${\bar K}^0$ being emitted from either particle 1 ($n$) or 2 ($n$).}
    \label{fig:00}
\end{figure*}

First, one needs to rewrite the Faddeev equations in terms of a different partition of diagrams. Following Refs.~\cite{Encarnacion:2025lyf,Ikeno:2025bsx,Agatao:2025ckp} we will consider partitions containing the contributions of the kaon interacting first with particle $i$ $(=1\text{ or }2)$ and ending with particle $j$ $(=1\text{ or }2)$. Note that, as we are working in the charge-particle basis and we consider exchange-type processes, the incoming and outgoing channels can be $K^-pn$, $K^-np$ and $\bar{K}^0nn$, named in short as d, e and $0$, respectively. Therefore the matrix elements will be denoted as $T^{\alpha \beta}_{ij}$, with $i,j \in \{1,2\}$ and $\alpha,\beta \in \{\text{d,e,0}\}$.
This means that we must construct a $6 \times 6$ matrix which has explicitly this form,
\begin{equation}
\mathbf{T} = \begin{pmatrix}
T^{\text{dd}}_{11} & T^{\text{dd}}_{12} & T^{\text{de}}_{11} &T^{\text{de}}_{12} &T^{\text{d0}}_{11} & T^{\text{d0}}_{12} \\ 
T^{\text{dd}}_{21} & T^{\text{dd}}_{22} & T^{\text{de}}_{21} &T^{\text{de}}_{22} &T^{\text{d0}}_{21} & T^{\text{d0}}_{22} \\ 
T^{\text{ed}}_{11} & T^{\text{ed}}_{12} & T^{\text{ee}}_{11} &T^{\text{ee}}_{12} &T^{\text{e0}}_{11} & T^{\text{e0}}_{12} \\ 
T^{\text{ed}}_{21} & T^{\text{ed}}_{22} & T^{\text{ee}}_{21} &T^{\text{ee}}_{22} &T^{\text{e0}}_{21} & T^{\text{e0}}_{22} \\ 
T^{\text{0d}}_{11} & T^{\text{0d}}_{12} & T^{\text{0e}}_{11} &T^{\text{0e}}_{12} &T^{\text{00}}_{11} & T^{\text{00}}_{12} \\ 
T^{\text{0d}}_{21} & T^{\text{0d}}_{22} & T^{\text{0e}}_{21} &T^{\text{0e}}_{22} &T^{\text{00}}_{21} & T^{\text{00}}_{22}  
\end{pmatrix} \ .
\label{eq:kernel}    
\end{equation}
The matrix elements are derived from the solution of a coupled set of equations that can be written from the diagrams represented in Figs.~\ref{fig:dd}-\ref{fig:00}.

The diagrams of Fig.~\ref{fig:dd} represent the equations for the $K^-pn \to K^- pn $ transition and the corresponding equations read,
\begin{eqnarray}
T^{\rm dd}_{11} &=& t_1 +t_1 G_-T^{\rm dd}_{21} + t_x G_0 T^{\rm 0d}_{21} \ , \label{eq:dd11}\\
T^{\rm dd}_{12} &=&  t_1 G_-T^{\rm dd}_{22} + t_x G_0 T^{\rm 0d}_{22} \ , \\
T^{\rm dd}_{21} &=&  t_2 G_-T^{\rm dd}_{11}  \ , \\
T^{\rm dd}_{22} &=&  t_2 + t_2 G_-T^{\rm dd}_{12} \ .
\end{eqnarray}

The diagrams of Fig.~\ref{fig:de} represent the equations for the $K^-pn \to K^- n p$ transition, where the $pn$ pair is exchanged in the final state, and the corresponding equations read,
\begin{eqnarray}
T^{\rm de}_{11} &=& t_1 G_-T^{\rm de}_{21} + t_x G_0 T^{\rm 0e}_{21} \ , \\
T^{\rm de}_{12} &=&  t_1 G_-T^{\rm de}_{22} + t_x G_0 T^{\rm 0e}_{22}  \ ,\\
T^{\rm de}_{21} &=&  t_2 G_-T^{\rm de}_{11} \ , \\
T^{\rm de}_{22} &=&  t_2 G_-T^{\rm de}_{12} \ .
\end{eqnarray}

The diagrams of Fig.~\ref{fig:0d} represent the equations for the ${\bar K}^0 nn \to K^- pn$ transition and the corresponding equations read,
\begin{eqnarray}
T^{\rm 0d}_{11} &=& t_x + t_0 G_0T^{\rm 0d}_{21} + t_x G_- T^{\rm dd}_{21} \ , \\
T^{\rm 0d}_{12} &=&  t_0 G_0T^{\rm 0d}_{22} + t_x G_- T^{\rm dd}_{22} \ , \\
T^{\rm 0d}_{21} &=&  t_0 G_0T^{\rm 0d}_{11} +  t_x G_- T^{\rm ed}_{11}  \ , \\
T^{\rm 0d}_{22} &=& t_0 G_0T^{\rm 0d}_{12} +  t_x G_-T^{\rm ed}_{12} \ . \label{eq:0d22}
\end{eqnarray}

Note that at this point we have 3 sets of 4 equations, but 16 unknowns because of the presence of the amplitudes 
$T^{\rm 0e}_{21}$, $T^{\rm 0e}_{22}$, $T^{\rm ed}_{11}$ and $T^{\rm ed}_{12}$. However, it can be easily checked topologically that the following symmetries are satisfied: $T^{\rm ee}= T^{\rm dd}(1\leftrightarrow 2)$, 
$T^{\rm ed}= T^{\rm de}(1\leftrightarrow 2)$, $T^{\rm 0e}= T^{\rm 0d}(1\leftrightarrow 2)$. These symmetries not only allow one to close the system of 12 equations with 12 unknowns, but also permit obtaining all the 24 matrix elements of the first 4 columns in the kernel matrix of Eq.~(\ref{eq:kernel}).

Next, we focus on the diagrams of Figs.~\ref{fig:d0} and \ref{fig:00}, which are clearly coupled, leading to the following system of eight equations:
\begin{eqnarray}
T^{\rm d0}_{11} &=& t_x + t_1 G_-T^{\rm d0}_{21} + t_x G_0 T^{\rm 00}_{21} \ , \\
T^{\rm d0}_{12} &=&  t_1 G_-T^{\rm d0}_{22} + t_x G_0 T^{\rm 00}_{22} \ , \\
T^{\rm d0}_{21} &=&   t_2 G_- T^{\rm d0}_{11} \ , \\
T^{\rm d0}_{22} &=&  t_2 G_-T^{\rm d0}_{12} \ , \\
T^{\rm 00}_{11} &=& t_0 + t_0 G_0T^{\rm 00}_{21} + t_x G_- T^{\rm d0}_{21} \ , \\
T^{\rm 00}_{12} &=&  t_0 G_0T^{\rm 00}_{22} + t_x G_- T^{\rm d0}_{22} \ , \\
T^{\rm 00}_{21} &=&  T^{\rm 00}_{12} \ , \\
T^{\rm 00}_{22} &=&  T^{\rm 00}_{11} \ .
\end{eqnarray}

Because of the last two identities, one formally needs to solve a coupled system of 6 linear equations with 6 unknowns, which, together with the following symmetries, $T^{\rm e0}= T^{\rm d0}(1\leftrightarrow 2)$, allow us to obtain the remaining elements of the last two columns of the kernel matrix of Eq.~(\ref{eq:kernel}). 

After solving the systems of equations, the following symmetries have additionally emerged:
$T^{\rm dd}_{12}=T^{\rm dd}_{21}$, $T^{\rm de}_{11}=T^{\rm de}_{22}$ and $T^{\rm 0d}_{ij}=T^{\rm d0}_{ji}$. With these considerations and keeping only the independent matrix elements, one realizes that the matrix $\mathbf{T}$ is symmetric:
\begin{equation}
\mathbf{T} = \begin{pmatrix}
T^{\text{dd}}_{11} & T^{\text{dd}}_{12} & T^{\text{de}}_{11} &T^{\text{de}}_{12} &T^{\text{d0}}_{11} & T^{\text{d0}}_{12} \\ 
T^{\text{dd}}_{12} & T^{\text{dd}}_{22} & T^{\text{de}}_{21} &T^{\text{de}}_{11} &T^{\text{d0}}_{21} & T^{\text{d0}}_{22} \\ 
T^{\text{de}}_{11} & T^{\text{de}}_{21} & T^{\text{dd}}_{22} &T^{\text{dd}}_{12} &T^{\text{d0}}_{22} & T^{\text{d0}}_{21} \\ 
T^{\text{de}}_{12} & T^{\text{de}}_{11} & T^{\text{dd}}_{12} &T^{\text{dd}}_{11} &T^{\text{d0}}_{12} & T^{\text{d0}}_{11} \\ 
T^{\text{d0}}_{11} & T^{\text{d0}}_{21} & T^{\text{d0}}_{22} &T^{\text{d0}}_{12} &T^{\text{00}}_{11} & T^{\text{00}}_{12} \\ 
T^{\text{d0}}_{12} & T^{\text{d0}}_{22} & T^{\text{d0}}_{21} &T^{\text{d0}}_{11} &T^{\text{00}}_{12} & T^{\text{00}}_{11}  
\end{pmatrix} \ .
\label{eq:kernel_sym}    
\end{equation}

While the explicit expressions for the matrix elements of $\mathbf{T}$ are lengthy, the ones of its inverse are found to be quite simple: 

\begin{widetext}
\begin{equation}
\mathbf{T}^{-1} = 
\begin{pmatrix}
\frac{t_0}{t_0 t_1 - t_x^2} & -G_- & 0 & 0 &  -\frac{t_x}{t_0 t_1 - t_x^2} & 0 \\  
-G_-& t_2^{-1} & 0 & 0& 0& 0 \\
0 & 0 & t_2^{-1} & -G_- & 0 & 0\\
0 & 0 & -G_- &  \frac{t_0}{t_0 t_1 - t_x^2} & 0 &-\frac{t_x}{t_0 t_1 - t_x^2} \\
-\frac{t_x}{t_0 t_1 - t_x^2} & 0 & 0 & 0 &  \frac{t_1}{t_0 t_1 - t_x^2} & -G_0 \\
0 & 0 & 0 & -\frac{t_x}{t_0 t_1 - t_x^2} & -G_0  &  \frac{t_1}{t_0 t_1 - t_x^2}
\end{pmatrix} 
\ . 
\label{eq:inverse_kernel}    
\end{equation}
\end{widetext}

Note that the above $\mathbf{T}^{-1}$ matrix would be reorganized in $2 \times 2$ boxes if $t_x\to 0$. Otherwise, all channels are coupled. 
The limit $t_x \to 0$ is physically motivated by the fact that the charge-exchange $K^- p \to {\bar K}^0 n$ process will transform the deuteron into a two-neutron cluster. Since the deuteron has total spin $J=1$ and the $s-$wave interaction with the kaon cannot change it, the final two-neutron system inherits this total spin value. Therefore, owing to antisymmetry arguments, the process controlled by $t_x$ cannot take place if the neutrons remain in a spatially-symmetric orbital wavefunction ($\ell=0,2$)\footnote{We thank Eulogio Oset for raising this point to us.}. However, since we have allowed the cluster to have a small recoil, the interaction with the kaon can provide a source of orbital angular momentum, allowing the two-neutron system to be in a relative $\ell=1$ state and the exchange process to be physically realized. Nevertheless, such a $|\Delta \ell|=1$ change is kinematically suppressed and, as we discuss below, the contribution of the exchange process is small. However, inelastic channels in particle-cluster scattering can be important in other systems, particularly those not subject to Pauli-exclusion constraints. Our aim is therefore to also establish a general formalism capable of accommodating such processes.

It is now instructive to show how the original expression of the FCA amplitude of Eq.~(\ref{eq:FCAfinal}) is recovered from the $\mathbf{T}$ matrix elements. This is done by realizing that $T_p$, $T_n$ and $T_x$ can be obtained from the following relations:
\begin{align} 
T_p &=T^{\text{dd}}_{11}+T^{\text{dd}}_{12} -T^{\text{de}}_{11}- T^{\text{de}}_{12} \ , \\
T_n & = T^{\text{dd}}_{21}+T^{\text{dd}}_{22} -T^{\text{de}}_{21}- T^{\text{de}}_{22} \ , \\ 
-T^\mathrm{x}_n  &= T^{\text 0d}_{21}+T^{\text{0d}}_{22} - T^{\text{0e}}_{21}- T^{\text{0e}}_{22} \nonumber \\ & -(T^{\text{0d}}_{11}+ T^{\text{0d}}_{12} -T^{\text{0d}}_{21}-T^{\text{0d}}_{22}) \ .
\end{align}

Then, employing Eqs.~(\ref{eq:dd11})-(\ref{eq:0d22}), we find

\begin{align}
T_p & = t_1 + t_1 G_- (T^{\text{dd}}_{21}+T^{\text{dd}}_{22} -T^{\text{de}}_{21}- T^{\text{de}}_{22}) \nonumber \\ 
& \,~~~~~~- t_x G_0 (T^{\text{0d}}_{11}+ T^{\text{0d}}_{12} -T^{\text{0d}}_{21}-T^{\text{0d}}_{22}) \nonumber \\
& = t_1 + t_1 G_- T_n -t_x G_0 T^\mathrm{x}_n  \ , \\
T_n & = t_2 + t_2 G_- ( T^{\text{dd}}_{11}+T^{\text{dd}}_{12} -T^{\text{de}}_{11}- T^{\text{de}}_{12}) \nonumber \\
&= t_2 +t_2 G_- T_p \ , \\
T^\mathrm{x}_n  & = t_x + t_0 G_0 (T^{\text 0d}_{21}+T^{\text{0d}}_{22} - T^{\text{0e}}_{21}-T^{\text{0e}}_{22}) \nonumber \\
& \,~~~~~~+ t_x G_- (T^{\text{dd}}_{21}+T^{\text{dd}}_{22} -T^{\text{de}}_{21}- T^{\text{de}}_{22}) \nonumber \\
& = t_x - t_0 G_0 T^\mathrm{x}_n + t_x G_- T_n \ .
\end{align}

These equations exactly coincide with Eq.~(\ref{eq:FCAnew}). We have also checked that by evaluating the combination
\begin{equation}
 T^{\rm FCA}_{K^- d}=T^p+T^n= \sum_{i,j=1}^2\left(T_{ij}^{\rm dd}-T_{ij}^{\rm de}\right) \ ,
 \end{equation}
employing the solutions for the required $T^{\alpha\beta}_{ij}$ amplitudes, we recover the FCA expression of Eq.~(\ref{eq:FCAfinal}) in terms of the two-body scattering amplitudes.

\subsubsection{Unitarity}

\begin{figure*}[htbp]
    \centering      \includegraphics[width=0.8\linewidth]{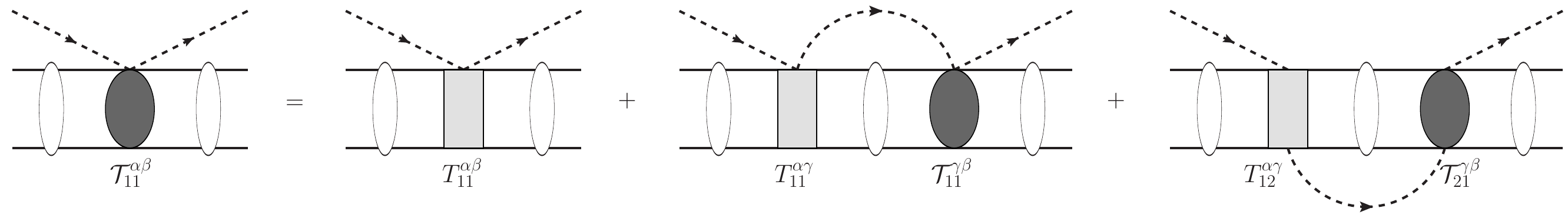}
    \caption{Diagrams involved in the coherent scattering of the $K^-$ with the deuteron cluster.}
    \label{fig:unit}
\end{figure*}

One of the drawbacks of the FCA is that it does not satisfy elastic unitarity, as emphasized in Ref.~\cite{Encarnacion:2025lyf,Ikeno:2025bsx,Agatao:2025ckp}. Indeed, close to a threshold, the effective range expansion of the inverse of the $T$ matrix should give (in our normalization),
\begin{equation}
    -\frac{4\pi\sqrt{s}}{M_d}T_{K^-d}^{-1} = \frac{1}{A_{K^-d}}+ \frac{1}{2} r_{K^- d}\ q_{\rm c.m.}^2 - {\rm i}\,q_{\rm c.m.} \ ,
    \label{eq:expansion}
\end{equation}
where $q_{\rm c.m.}$ is the momentum of the $K^-$ in the $K^-d$ center-of-mass system. In addition to the factor $(-{\rm i})$, the linear term in $q_{\rm c.m.}$ of this expansion should carry a coefficient of 1. Instead, in the FCA, this coefficient is found to be:
\begin{widetext}
\begin{equation}
\frac{({t_2}\Re G +1) \left(2   {t_2}
    t_x^4\Re G^3-t_x^2 ( {t_2}\Re G (4 {t_1}\text {Re}\,G 
   ( {t_0}\Re G+1)+2  {t_0} \Re G+3)+1)+2 {t_1}
   {t_2} ( {t_0}\Re G + 1)^2 (
   {t_1}\Re G +1)\right)}{\left(( {t_0}\Re G + 1) (2 
   {t_1} {t_2}\Re G+{t_1}+{t_2})-t_x^2 \Re G ( {t_2} 2
   \Re G + 1)\right)^2} \ ,
   \label{eq:coeff}
   \end{equation}
\end{widetext}
where in the above expression, we have assumed $G_0 = G_{-}\equiv G$, for simplicity,  and $\Re G$ stands for the real part of $G$.

More particularly, in the absence of the exchange term $t_x$ the linear coefficient in the FCA is 
\begin{eqnarray}
&&\frac{
2 \,t_1 t_2 (1 + t_1 \,\Re G_0 ) (1 + t_2 \,\Re G_0)}{(t_1 + t_2 + 2\, t_1 t_2 \,\Re G_0))^2}  \nonumber \\
&=&
\frac{1}{2}-\frac{1}{2}\frac{(t_1-t_2)^2}{(t_1+t_2+2\,t_1 t_2  \,\Re G_0)^2} \ ,
\end{eqnarray}
which coincides with the expression of Eq.~(17) in Ref.~\cite{Encarnacion:2025lyf}.

It is clear that the FCA approximation violates elastic unitarity. The degree of this violation for the $K^-d$ amplitude depends on how close to unity the coefficient in Eq.~(\ref{eq:coeff}) is. In our previous work~\cite{Ramos:2025ibe}, where we employed the FCA, we matched the opening of the two-body $K^- d$ threshold to that of the three-body $K^-pn$ channel. This effectively caused the unitary cut to open at the $K^- d$ threshold already at the impulse-approximation level. In that case, the coefficient in Eq.~(\ref{eq:coeff}) was found to be about $0.7$, relatively close to unity.

In Ref.~\cite{Ikeno:2025bsx}, a procedure for implementing elastic unitarity by incorporating the coherent scattering of the particle from the cluster was proposed and subsequently mathematically justified in Ref.~\cite{Agatao:2025ckp}. The procedure was formulated using isospin-averaged amplitudes and neglecting exchange terms. In the present work, as mentioned above, we extend this procedure to the $K^-d$ system by explicitly accounting for the exchange terms. Following the procedure of Refs.~\cite{Ikeno:2025bsx,Agatao:2025ckp} we implement the coherent scattering of the $K^-$ from the deuteron cluster by resumming an infinite series of diagrams. In Fig.~\ref{fig:unit} we display these diagrams only for the ``11" element of the $\beta\to\alpha$ transition unitary amplitude, ${\cal T}^{\alpha\beta}_{11}$. 
Similar plots can be done for all the other matrix elements, ${\cal T}_{ij}^{\alpha\beta}$. Together, they form the following coupled-channel set of equations:
\begin{equation}
    {\cal T}^{\alpha\beta}_{ij}={T}^{\alpha\beta}_{ij} + {T}^{\alpha\gamma}_{ik} G_{c}^{k}  {\cal T}^{\gamma\beta}_{kj}
\end{equation}
or, in matrix form:
\begin{equation}
\mathcal{T} = \mathbf{T} + \mathbf{T} \, \mathbf{G}_c \,\mathcal{T}
\label{eq:unit_ampl}
\end{equation}
where  $\mathbf{T}$ is the symmetric kernel of Eq.~(\ref{eq:kernel_sym}) and $\mathbf{G}_c=\text{diag}(G_-^{(1)},G_-^{(2)},G_-^{(1)},G_-^{(2)},G_0^{(1)},G_0^{(2)})$ is a diagonal matrix containing the $K^-(pn)$ propagators ($G_-^{(1)}$ or $G_-^{(2)}$) and ${\bar K}^0(nn)$ propagators ($G_0^{(1)}$ or $G_0^{(2)}$), for the kaon being emitted and absorbed in particle 1 or 2 of the cluster, respectively.
These propagators are given by
\begin{equation}
    G_{\rm m}^{(i)}= \int\frac{\mathrm{d}^3 q}{(2\pi)^3}\frac{\left(F_d^{(i)}(\bm{q})\right)^2}{2\omega_{\rm m}(\bm{q})\left(\sqrt{s_{Kd}}-\omega_{\rm m}(\bm{q})-E_{\rm m}(\bm{q})+{\rm i}\eta \right) } \ ,
        \label{eq:propcluster}
\end{equation}
with $i=1,2$, and contain a deuteron form-factor squared, one for each vertex involving a virtual kaon, as explained in Ref.~\cite{Ikeno:2025bsx}. The argument of the vertex form factor for a cluster composed of constituents with unequal masses, such as the proton and neutron in the present case, is derived in Ref.~\cite{Yamagata-Sekihara:2010kpd}. Assuming a small external kaon momentum, this argument can be approximated as
$\bm{q}\,m_2/(m_1 + m_2)$ or $\bm{q}\,m_1/(m_1 + m_2)$ for $G_{\rm m}^{(1)}$ or $G_{\rm m}^{(2)}$, respectively, as done in Ref.~\cite{Ikeno:2025bsx}.

An alternative way to write Eq.~(\ref{eq:unit_ampl}) is
\begin{equation}
\mathcal{T}^{-1} = \mathbf{T}^{-1} - \mathbf{G}_c \ ,
\label{eq:unit_ampl_2}
\end{equation}
where $\mathbf{T}^{-1}$ is given by Eq.~\eqref{eq:inverse_kernel}.

Once Eq.~(\ref{eq:unit_ampl}), or Eq.~(\ref{eq:unit_ampl_2}), is solved the final unitarized $K^- d$ amplitude is obtained from
\begin{equation}
 T_{K^- d}^{\rm uni}=\sum_{i,j=1}^2\left(\mathcal{T}^{\rm dd}_{ij}-\mathcal{T}^{\rm de}_{ij}\right) \ .
 \end{equation}

The corresponding analytic expression in terms of elementary amplitudes and propagators is lengthy and not reported here.  But we comment on some interesting limits:

\begin{enumerate} 
\item If we set all propagators to be the same, namely $G_-^{(1)}=G_-^{(2)}=G_-$ and $G_0^{(1)}=G_0^{(2)}=G_0$, we obtain
\begin{equation}
 T^{K^- d}_{\rm uni} \to 
\frac{t_1+t_2}{1-G_-(t_1+t_2)} \ ,
\end{equation}
or, equivalently:
\begin{equation}
\left(T^{K^- d}_{\rm uni}\right)^{-1}  \to \frac{1}{t_1+t_2}-G_-
\end{equation}
Note that, in this case, $ T^{K^- d}_{\rm uni}$ does not depend on $t_x$ (nor on $G_0$) and it is simply obtained from the solution of the unitarization equation employing the impulse approximation kernel (built from sum of the two elementary amplitudes $t_1$ and $t_2$).

\item 
Going back to the full length expression of $T^{K^- d}_{\rm uni}$ and taking the limit $t_x\to 0$, which also implies the disappearance of all $G_0$, $G_0^{(1)}$ and  $G_0^{(2)}$ propagators, we obtain:
\begin{equation} 
T^{K^- d}_{\rm uni} \to \frac{{ t_1}+ { t_2}+\left(2  {G_-}-  {G_-^{(1)}}- {G_-^{(2)}}\right) { t_1}  { t_2}  }{ 1 - {G_-^{(1)}}  { t_1} - {G_-^{(2)}}  { t_2} - \left(
G_-^2 -  {G_-^{(1)}} {G_-^{(2)}}\right) t_1 t_2  } \ ,
\end{equation} 
hence recovering the expression given in the work of Ref.~\cite{Agatao:2025ckp}, where exchange mechanisms were not considered.
\end{enumerate}

\section{Results~\label{sec:results}}

\subsection{\mbox{\boldmath $K^-d$} low-energy observables}

In Table~\ref{tab:low_en} we present the calculated low-energy parameters of the $T_{K^-d}$ scattering amplitude, namely the scattering length $A_{K^-d}$, the effective range $r_{K^-d}$, and the coefficient accompanying the $-{\rm i}q_{\rm c.m.}$ term in the expansion of Eq.~(\ref{eq:expansion}), for the two interaction models (OR and BCN) and the different approximations considered in this work. 

    \begin{table}[ht!]
        \centering
        \begin{tabular}{l l c c c}
        \hline
        \hline
           Model & Approx & $A_{K^-d}\, [\mathrm{fm}]$ & $r_{K^-d}\, [\mathrm{fm}]$  & coeff. \\           
        \hline
 \multirow{3}{*}{OR} &
            FCA \cite{Ramos:2025ibe} &  $-2.06+{\rm i} 1.77$ & -  & 0.69 \\
 &  FCA & $-2.00+{\rm i} 2.23$ & -  & 0.38 \\
         &   UNI 
         & $-1.04+{\rm i} 0.90$ &
              $2.27 -{\rm i}0.71$ & 1.00 \\           
        & UNI$^{(*)}$ 
         &    $-1.12+{\rm i} 0.84$ &
         $1.78 -{\rm i}1.16$ &    1.00\\
                      \hline
  \multirow{3}{*}{BCN} &
            FCA \cite{Ramos:2025ibe}  &  $-1.82+{\rm i} 1.75$ & - & 0.62\\
          &  FCA & $-1.53+{\rm i} 2.04$ & -  & 0.40 \\
%
    &   UNI 
         & $-0.85+{\rm i} 1.07$ &
              $2.36 -{\rm i}0.72$ & 1.00\\              
              &                UNI$^{(*)}$ 
         &   $-0.93+{\rm i} 1.00$ &
 $2.04 -{\rm i}1.31$ &    1.00\\ 
                       \hline
        EXP\cite{ALICE:2026pxr} & \multicolumn{4}{l}{$\Re(A_{K^-d}) = -1.44\pm 0.15\text{(stat)}^{+0.10}_{-0.10}\text{(syst)}$ fm}\\ 
       & \multicolumn{4}{l}{$\Im(A_{K^-d}) = 1.34\pm 0.33\text{(stat)}^{+0.21}_{-0.15}\text{(syst)}$ fm} \\ 
        \hline
        \hline
        \end{tabular}
                \caption{Low energy scattering parameters for $K^-d$.\\
                $^{(*)}$ result when removing the exchange terms ($t_x\to 0$).}
        \label{tab:low_en}
    \end{table}

For completeness, we also include, in the first row for each model, the FCA results obtained in Ref.~\cite{Ramos:2025ibe}. There, unitarity was not explicitly implemented; instead, its effects were approximately compensated by shifting the FCA amplitude by $2.22$ MeV, corresponding to the deuteron binding energy, so that the two-body $K^-d$ threshold coincided with that of the three-body $K^-pn$ system. As shown in Table~\ref{tab:low_en} this prescription yields a FCA ``unitary" coefficient from Eq.~(\ref{eq:coeff}) of $0.6-0.7$, closer to unity than the value of about $0.4$ obtained with the current unshifted FCA amplitude. It is also worth noting that the energy shift applied to the FCA amplitudes in Ref.~\cite{Ramos:2025ibe} induces a sizable change in the $K^-d$ scattering length.

Once the coherent scattering of the $K^-$ from the deuteron is implemented, exact unitarity is restored, with the coefficient of the linear $-{\rm i}q_{\rm c.m.}$ term in the low-energy expansion of the amplitude [Eq.~(\ref{eq:expansion})] taking the value one. Since the low-energy expansion is reliable in this case, the effective range $r_{K^-d}$ can be extracted from the coefficient of the quadratic term in $q_{\rm c.m.}$ and its value is also reported in Table~\ref{tab:low_en}.

We observe that the unitary value of $A_{K^-d}$ differs substantially from the corresponding FCA result, being reduced by a factor close to two in both OR and BCN models. This sizable variation is related  to the existence of a subthreshold resonant structure in the $K^-d$ scattering amplitude which is tied to the $\Lambda(1405)$ resonance governing the elementary ${\bar K}N$ $t$-matrices. The coherent propagation acts as a binding mechanism, shifting the resonant structure towards lower energies further away from the $K^-d$ threshold. Consequently, the amplitude at this energy is strongly modified, acquiring a more moderate value.

It is interesting to see that the coherent propagation exerts the same qualitative effect in both OR and BCN models. Upon comparing our final results with the $K^- d$ scattering length derived recently by the ALICE collaboration \cite{ALICE:2026pxr}, we observe that both models can reproduce the imaginary part within errors but not the real part. We recall that the experimental $K^- d$ scattering is not a direct measurement, but it is in fact extracted from fits to correlation functions measured by the ALICE collaboration employing the Lednický–Lyuboshitz (LL) model \cite{Lednicky:1981su}, which only accounts for the asymptotic behavior of the wave function. The LL correlation function was shown to differ significantly from that obtained with a wave-function having the proper short-range behavior \cite{Ramos:2025ibe}, thus a future improved analysis that takes this aspect into consideration is necessary.

We finally comment on the results obtained when we set $t_x\to 0$, labeled with UNI$^{(*)}$. The scattering length experiences a moderate modification of up to 10\% with respect to the result which includes the charge exchange $K^- d \to {\bar K}^0 (nn)$ transition. This is a numerical proof of the arguments given above on the antisymmetry constraints, which in practice suppress the effect of the exchange processes in the $K^-d$ system. However, the influence of the exchange terms in the scattering range is more noticeable because this observable is tied to the amplitude at higher energies where the recoil effects, responsible for conveying orbital momentum to the intermediate $(nn)$ pair, are stronger.

    \begin{table}[h!]
        \centering
        \begin{tabular}{l l c c c c }
        \hline
        \hline
           Model & Approx & 
           \multicolumn{2}{c}{Corrected DT} & \multicolumn{2}{c}{Summed-up DT} \\
           &  & \!\!$\varepsilon_{{\rm 1s}}^{d}$ $[\mathrm{eV}]$\!\! & \!\!$\Gamma_{{\rm 1s}}^{d}$ $[\mathrm{eV}]$ &
            $\varepsilon_{{\rm 1s}}^{d}$  $[\mathrm{eV}]$\!\! & \!\!$\Gamma_{{\rm 1s}}^{d}$  $[\mathrm{eV}]$ 
           \\           
        \hline
 \multirow{3}{*}{OR} &
            FCA \cite{Ramos:2025ibe} & 1124 &  626 & 1040 & 1106 \\
        &  FCA & 1304 & 849 & 1092 & 1379 \\   
          &   UNI & 600 & 696 & 583 & 767  \\
          &  UNI$^{(*)}$ &   617 &  623  &  607 &  701 \\          
                      \hline
  \multirow{3}{*}{BCN} &
            FCA \cite{Ramos:2025ibe} & 1069 &  793 & 968 & 1162 \\
          &  FCA & 1110 & 1172 & 928 & 1435 \\
         &   UNI & 557 & 913 & 525 & 958 \\ 
          &    UNI$^{(*)}$ &   574 & 820  &   547 &  876 \\ 
         \hline
        EXP~\cite{Bazzi:2026gri} & \multicolumn{5}{l}{$\varepsilon_{{\rm 1s}}^{d} = 810.9\pm 24.5\text{(stat)}\pm 2.1\text{(syst)}$ eV }\\ 
       & \multicolumn{5}{l}{$\Gamma_{{\rm 1s}}^{d} = 812\pm 97\text{(stat)}\pm 33\text{(syst)}$ eV} \\        
        \hline
        \hline
        \end{tabular}
                \caption{Energy shift and width of the 1s $K^-d$ atomic level \\
                $^{(*)}$ result when removing the exchange terms ($t_x\to 0$).}
        \label{tab:shift_width}
    \end{table}

In Table~\ref{tab:shift_width} we display our results for the shift, $\varepsilon_{{\rm 1s}}^{d}$, and width, $\Gamma_{{\rm 1s}}^{d}$, of the 1s level of kaonic deuterium, obtained from the corrected Deser-Trueman formula~\cite{Meissner:2005bz}
\begin{equation}
\varepsilon^d_{1s} - {\rm i}\frac{\Gamma^d_{1s}}{2} =
- 2 \alpha^3 \mu_r^2 A_{K^- d} \left\{1 -2\alpha \mu_r A_{K^- d} (\ln\alpha -1)\right\} \ ,
\label{eq:corrected}
\end{equation}
or the summed-up one \cite{Baru:2009tx,Shevchenko:2021swf} 
\begin{equation}
\varepsilon^d_{1s} - {\rm i}\frac{\Gamma^d_{1s}}{2} =
- \frac{2 \alpha^3 \mu_r^2 A_{K^- d}}{1 +2\alpha \mu_r A_{K^- d} (\ln\alpha -1)}  \ ,
\label{eq:summedup}
\end{equation}
for the two interaction models and approximations employed in the present work. 
In these equations $\alpha$ is the fine-structure constant and $\mu_r$ is the kaon-deuteron reduced mass.

The FCA scheme of Ref.~\cite{Ramos:2025ibe} delivers, for both OR and BCN models, a large energy shift $\varepsilon_{1s}^d$ alongside a moderate 1s width $\Gamma_{1s}^d$ with the corrected Deser-Trueman formula [Eq.~(\ref{eq:corrected})]. The summed-up expression [Eq.~(\ref{eq:summedup})] slightly reduces the shift and significantly enhances the width, and both observables remain outside the experimental error band~\cite{Bazzi:2026gri}.

Within the present FCA framework, which accounts for the nucleon binding in the elementary ${\bar K}N$ amplitudes [Eqs.~(\ref{eq:two-body-energy1}) and (\ref{eq:two-body-energy2})], the corrected Deser-Trueman formula gives a substantially larger width that gets even more enhanced by the summed-up expression and departs significantly from the experimental result. Nevertheless, the reliability of this FCA scheme is questionable as it violates unitarity drastically.

More interesting are the results obtained with the coherent scattering approach that preserves unitarity. In this case, and for both OR and BCN models, the energy shift and the width acquire more moderate, balanced values than the FCA within the corrected Deser-Trueman formula. 
This is directly connected to the smaller values of the scattering length obtained within the unitary procedure. Consequently, the higher-order corrections implemented by the summed-up expression have a noticeably weaker influence.  Within this prescription, which has been shown to provide a quite accurate approximation to the exact solution of the
Schr\"odinger equation \cite{Shevchenko:2021swf}, the energy shift of the two models fails to reproduce the precise experimental value, while the OR width is compatible with experiment and the BCN one falls shortly away from the upper side of the experimental error.  As in the case of the scattering length, neglecting charge-exchange processes leads to a moderate change in the 1s kaonic-deuterium observables: the energy shift increases by about 4\%, while the width decreases by roughly 9\%.

\subsection{Constraints from SIDDHARTA-2 kaonic atoms}

We present in this section a preliminary study aimed at assessing the potential constraining power of the SIDDHARTA-2 measurements of the shift and width of the 1s-state in kaonic deuterium \cite{Bazzi:2026gri} and hydrogen \cite{Bazzi:2026zmc} on coupled-channel, chirally motivated ${\bar K}N$ interaction models.
We focus on the BCN model, which, we recall, is derived  from a chiral $SU(3)$ Lagrangian up to NLO and implements coupled-channel unitarization. The resulting interaction has $16$ parameters, $p_i^{\rm BCN}$, $i=1,\ldots,16$, comprising low-energy constants and regularization-related subtraction constants. As mentioned above, the parameters are constrained by a large set of experimental data, including threshold observables as well as elastic and inelastic two-body $\bar{K}N$ transition observables. Among these threshold observables, the shift and width of the 1s kaonic-hydrogen level from the first SIDDHARTA measurement~\cite{SIDDHARTA:2011dsy}---novel at the time---were employed. These values are shown in Table~\ref{tab:shift_width_hidrogen}, together with those of the more recent and precise SIDDHARTA-2 measurement \cite{Bazzi:2026zmc} and the corresponding BCN theoretical output. The fit also provides the corresponding uncertainties, $\delta p_i^{\rm BCN}$, for the fitted parameter values. Both the parameter values and their uncertainties are reported in the first column of Table~II of Ref.~\cite{Feijoo:2018den}.

    \begin{table}[t]
        \centering
        \begin{tabular}{l c c c c }
        \hline
        \hline
           Model &  
           \multicolumn{2}{c}{Corrected DT} & \multicolumn{2}{c}{Summed-up DT} \\
           & \!\!$\varepsilon_{{\rm 1s}}$  $[\mathrm{eV}]$\!\! & \!\!$\Gamma_{{\rm 1s}}$  $[\mathrm{eV}]$ &
            $\varepsilon_{{\rm 1s}}$  $[\mathrm{eV}]$\!\! & \!\!$\Gamma_{{\rm 1s}}$ $[\mathrm{eV}]$ 
           \\           
        \hline
         BCN  &
             294 &  578 & 285 & 587 \\
         \hline
        SIDD~\cite{SIDDHARTA:2011dsy} & \multicolumn{4}{l}{$\varepsilon_{{\rm 1s}} = 283\pm 36\text{(stat)}\pm 6\text{(syst)}$ eV }\\ 
       & \multicolumn{4}{l}{$\Gamma_{{\rm 1s}} = 541\pm 89\text{(stat)}\pm 22\text{(syst)}$ eV} \\     
       SIDD2~\cite{Bazzi:2026zmc} & \multicolumn{4}{l}{$\varepsilon_{{\rm 1s}} = 303\pm 17\text{(stat)}\pm 2.5\text{(syst)}$ eV }\\ 
       & \multicolumn{4}{l}{$\Gamma_{{\rm 1s}} = 607\pm 62\text{(stat)}\pm 6\text{(syst)}$ eV} \\ 
        \hline
        \hline
        \end{tabular}
      \caption{Energy shift and width of the 1s kaonic hydrogen atomic level obtained using the BCN model, together with the corresponding experimental results from SIDDHARTA and SIDDHARTA-2.}
        \label{tab:shift_width_hidrogen}
    \end{table}
It is important to stress that we are not presenting here a standard compatibility study between the measured shifts and widths and the model predictions. In such an analysis, one would typically perform a Monte Carlo sampling of the model parameter space, preserving the correlations among the parameters through the covariance matrix obtained from the original fit, and determine whether the SIDDHARTA-2 measurements lie within the corresponding predicted distributions of $\varepsilon_{{\rm 1s}}$, $\Gamma_{{\rm 1s}}$, $\varepsilon^d_{{\rm 1s}}$ and $\Gamma^d_{{\rm 1s}}$.

Instead, we investigate to what extent the parameters of the BCN model can be fine-tuned to accommodate the newly measured kaonic-deuterium and kaonic-hydrogen energy shift and width, and whether these observables can help lift parameter degeneracies associated with correlations in the original fit. To this end, we generate $10{,}000$ pseudo-data sets for the kaonic-deuterium and kaonic-hydrogen 1s-state energy shift and width, sampling around the central values according to the corresponding experimental uncertainties reported by SIDDHARTA-2~\cite{Bazzi:2026gri,Bazzi:2026zmc}. These sets constitute the input for a bootstrap analysis, in which the model is refitted to each pseudo-data set. 

The simultaneous use of kaonic-hydrogen and kaonic-deuterium observables is particularly relevant because the two systems probe complementary combinations of the underlying $\bar KN$ amplitudes. In particular, the kaonic-hydrogen observables are primarily sensitive to the $K^-p$ interaction, whereas kaonic deuterium provides additional sensitivity to the $K^-n$ amplitude. Since the proton and neutron amplitudes involve different combinations of the isospin components of the $\bar KN$ interaction, the combined information from the two kaonic atoms provides sensitivity to directions in the parameter space that are less constrained by kaonic-hydrogen data alone. This complementarity is especially relevant for the isovector component, whose contribution is comparatively poorly constrained by existing data.

To explore the flexibility of the BCN parameterization in accommodating the new kaonic-deuterium and kaonic-hydrogen data, we define the following objective function:
\begin{widetext}
\begin{equation}
\chi^2 =
\frac{(\varepsilon_{{\rm 1s}}-\varepsilon_{{\rm 1s}}^{\rm exp})^2}
{(\delta\varepsilon_{{\rm 1s}}^{\rm exp})^2}
+
\frac{(\Gamma_{{\rm 1s}}-\Gamma_{{\rm 1s}}^{\rm exp})^2}
{(\delta\Gamma_{{\rm 1s}}^{\rm exp})^2}
+
\frac{(\varepsilon^d_{{\rm 1s}}-\varepsilon_{{\rm 1s}}^{d,\rm exp})^2}
{(\delta\varepsilon_{{\rm 1s}}^{d,\rm exp})^2}
+
\frac{(\Gamma^d_{{\rm 1s}}-\Gamma_{{\rm 1s}}^{d,\rm exp})^2}
{(\delta\Gamma_{{\rm 1s}}^{d,\rm exp})^2}
+
\sum_{i=1}^{16}
\frac{(p_i-p_i^{\rm BCN})^2}
{(\delta p_i^{\rm BCN})^2} \, .
\end{equation}
\end{widetext}

The last term on the right-hand side retains information from the BCN fit by penalizing deviations of the individual parameters from their respective original values. These priors therefore allow us to investigate whether the new light kaonic-atom observables can be accommodated within the parameter ranges allowed by the BCN fit, without imposing the correlations between the parameters encoded in the original covariance matrix. 

It should be noted, however, that an inherent bias is introduced by the prescription adopted to compute the energy shift and width of the \(1s\) level of the kaonic atoms under consideration. In the BCN fit, the corrected Deser-Trueman formula was employed [Eq.~(\ref{eq:corrected})], making this prescription the consistent choice for the present analysis. We nevertheless perform a parallel analysis using the summed-up Deser-Trueman formula [Eq.~(\ref{eq:summedup})], not merely for completeness, but also to assess the sensitivity of the results to the treatment of electromagnetic corrections. As will be shown below, the differences between the two prescriptions are relatively minor.

Within this framework, the outcome of the bootstrap analysis provides a test of the flexibility of the original BCN parameterization when confronted with the new kaonic-atom measurements. If the measured shifts and widths cannot be reproduced within their experimental uncertainties, even after allowing the individual parameters to vary within the ranges defined by the priors, this would point to a potential tension between the SIDDHARTA-2 measurements and the original BCN parameterization. In such a case, a global refit incorporating the kaonic-deuterium and kaonic-hydrogen data, together with the scattering and other low-energy observables used in the original analysis, would be required to determine whether the tension can be resolved through a correlated readjustment of the model parameters. Conversely, if the new observables can be accommodated within the allowed parameter ranges, this would indicate that the existing parameterization has sufficient flexibility to describe the new measurements without requiring a substantial modification of the underlying interaction. 

\begin{figure}[h]
    \centering
    \includegraphics[width=\columnwidth]{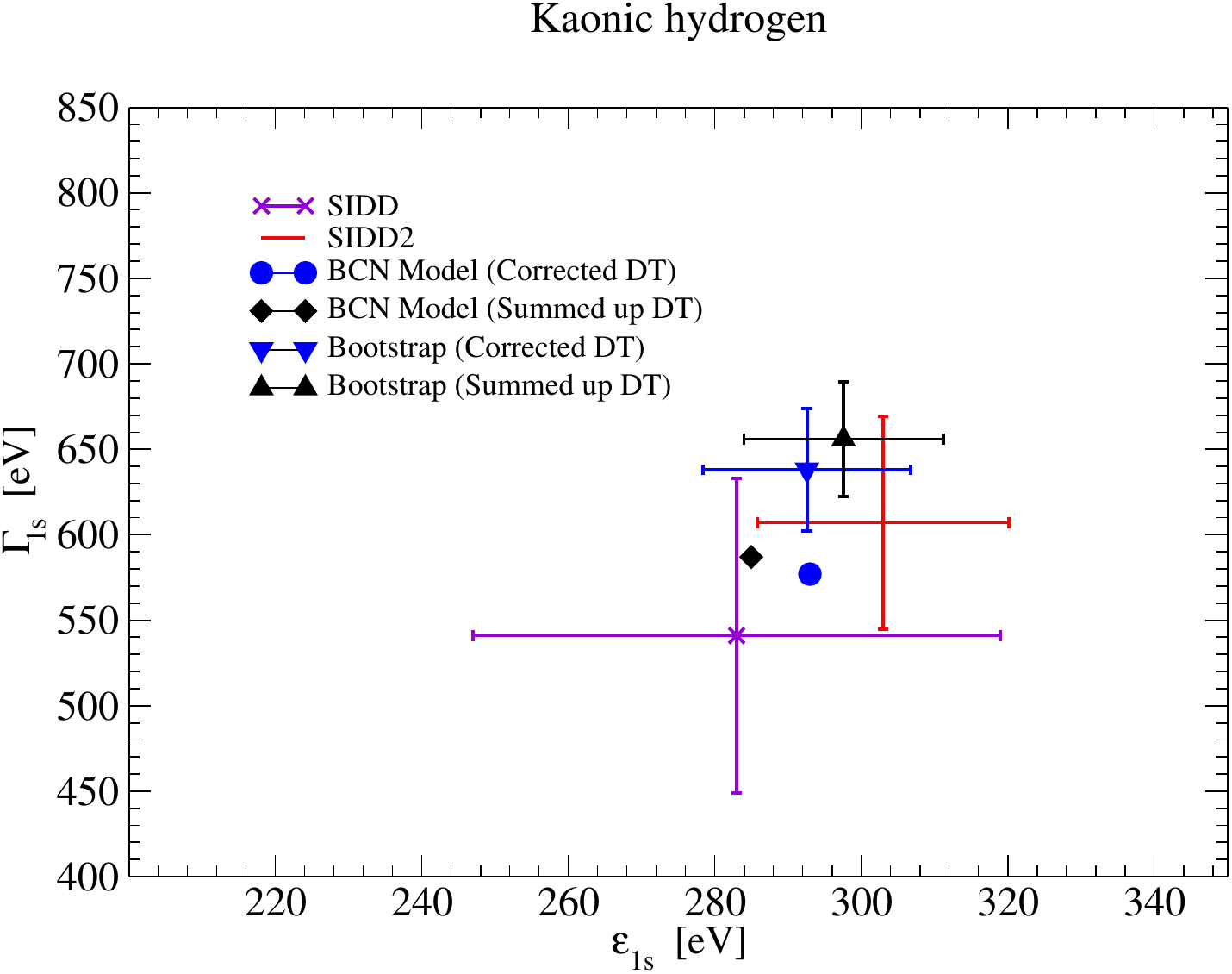}\\[0.5em]
    \includegraphics[width=\columnwidth]{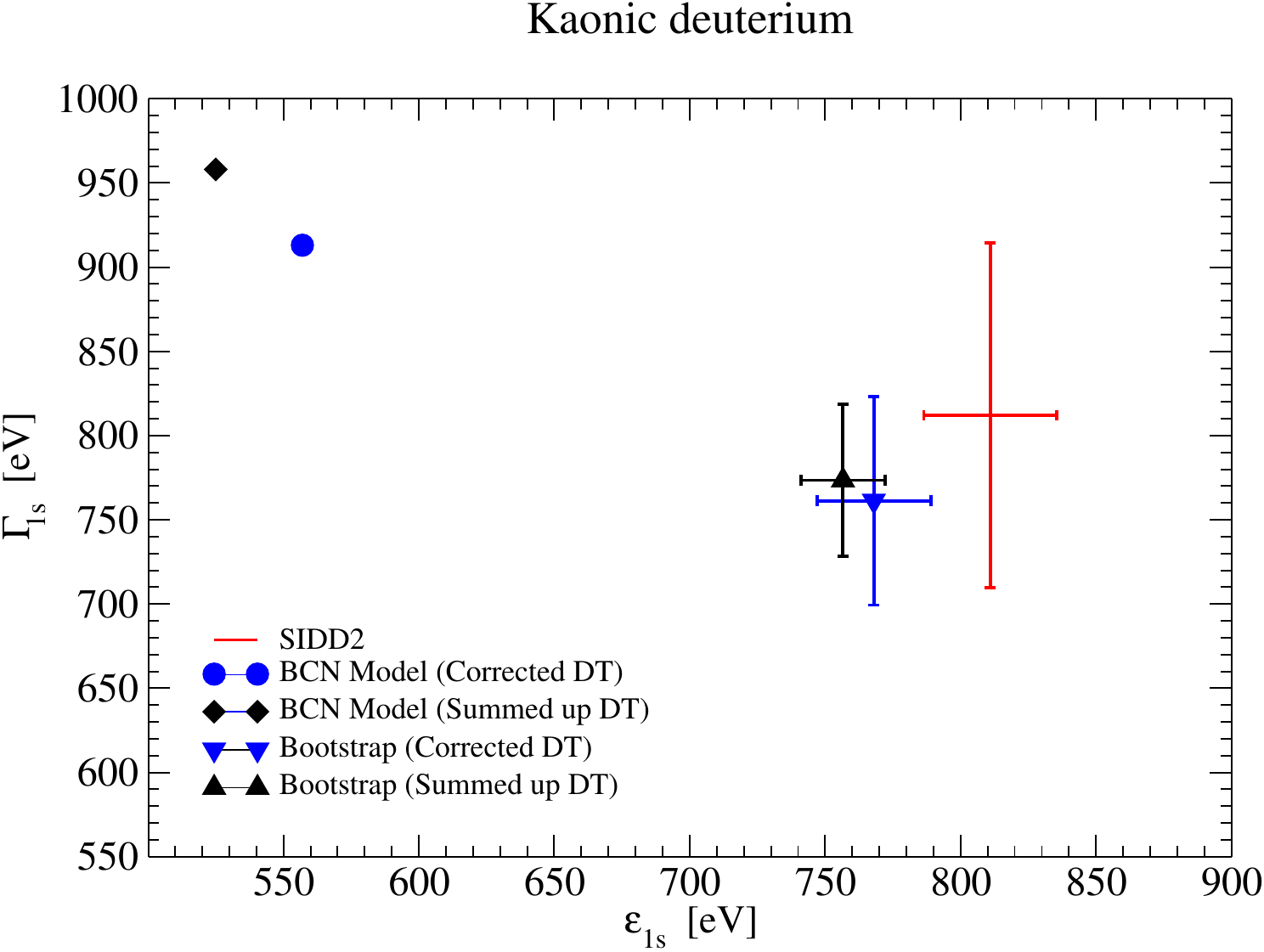}
\caption{Energy shift and width of the 1s atomic level obtained from the bootstrap analysis for the $K^-p$ (top panel) and $K^-d$ (bottom panel) systems, together with the corresponding values obtained using the BCN parameterization (diamonds and circles). The results obtained with the corrected  (blue markers) and summed-up (black markers) Deser-Trueman prescription are shown, together with the corresponding experimental SIDDHARTA-2 measurements~\cite{Bazzi:2026gri,Bazzi:2026zmc} (red crosses). For the $K^-p$ system, the previous SIDDHARTA measurement~\cite{SIDDHARTA:2011dsy} is also included (violet cross).}
    \label{fig:Katoms}
\end{figure}

In Table~\ref{tab:shift_width_bootstrap} we present the 
 energy shift and width of the 1s atomic level of kaonic hydrogen and kaonic deuterium obtained from the bootstrap analysis, together with their corresponding uncertainties.
    \begin{table}[H]
        \centering
        \begin{tabular}{l c c c c }
        \hline
        \hline
           Kaonic atom &  
           \multicolumn{2}{c}{Corrected DT} & \multicolumn{2}{c}{Summed-up DT} \\
           & \!\!$\varepsilon_{{\rm 1s}}$  $[\mathrm{eV}]$\!\! & \!\!$\Gamma_{{\rm 1s}}$  $[\mathrm{eV}]$ &
            $\varepsilon_{{\rm 1s}}$  $[\mathrm{eV}]$\!\! & \!\!$\Gamma_{{\rm 1s}}$ $[\mathrm{eV}]$ 
           \\           
        \hline
        $K^-p$ & $ 293\pm 14$ &  $ 638\pm 36$  & $ 297\pm 14$ & $ 656\pm 33$ \\
        \hline
         $K^-d$  & $ 768 \pm 21 $ &  $ 761 \pm 62 $ & $ 756\pm 16$ & $ 773\pm 45$ \\
        \hline
        \hline
        \end{tabular}
      \caption{Energy shift and width of the 1s kaonic-hydrogen and kaonic-deuterium atomic levels obtained from the bootstrap analysis of the BCN chiral model.}
        \label{tab:shift_width_bootstrap}
    \end{table}
 To facilitate the visualization, these results are shown in the top and bottom panels of Fig.~\ref{fig:Katoms} for kaonic hydrogen and deuterium, respectively, together with the experimental SIDDHARTA~\cite{SIDDHARTA:2011dsy} and SIDDHARTA-2~\cite{Bazzi:2026gri,Bazzi:2026zmc} measurements and the predictions of the original BCN model for both Deser-Trueman prescriptions (blue circles and black diamonds). When the corrected Deser-Trueman prescription is adopted, the bootstrap results for the kaonic-hidrogen and kaonic-deuterium observables (blue inverted triangles) are found to be compatible with the experimental data. When the summed-up Deser-Trueman formula is employed (black triangles), the $K^-p$ shift and width remain in good agreement with the corresponding experimental values, whereas the uncertainty intervals of the bootstrap analysis for the $K^-d$ observables do not overlap with those of the experimental measurements, although their respective boundaries are very close. 
The comparison between the two Deser-Trueman prescriptions should, however, be interpreted keeping in mind the 
 aforementioned bias of the original BCN parameterization which employed the corrected Deser-Trueman formula:
 differences between the two treatments, even if systematic, do not by themselves establish an improvement in the description, but rather provide an indication of the role played by the electromagnetic corrections. The small discrepancies observed in the results of Table~~\ref{tab:shift_width_bootstrap} indicate that the higher-order electromagnetic corrections incorporated in the summed-up Deser-Trueman treatment do affect the predicted observables, although the effect is relatively minor. 

\begin{figure*}[t]
    \centering

    \includegraphics[width=0.39\textwidth]{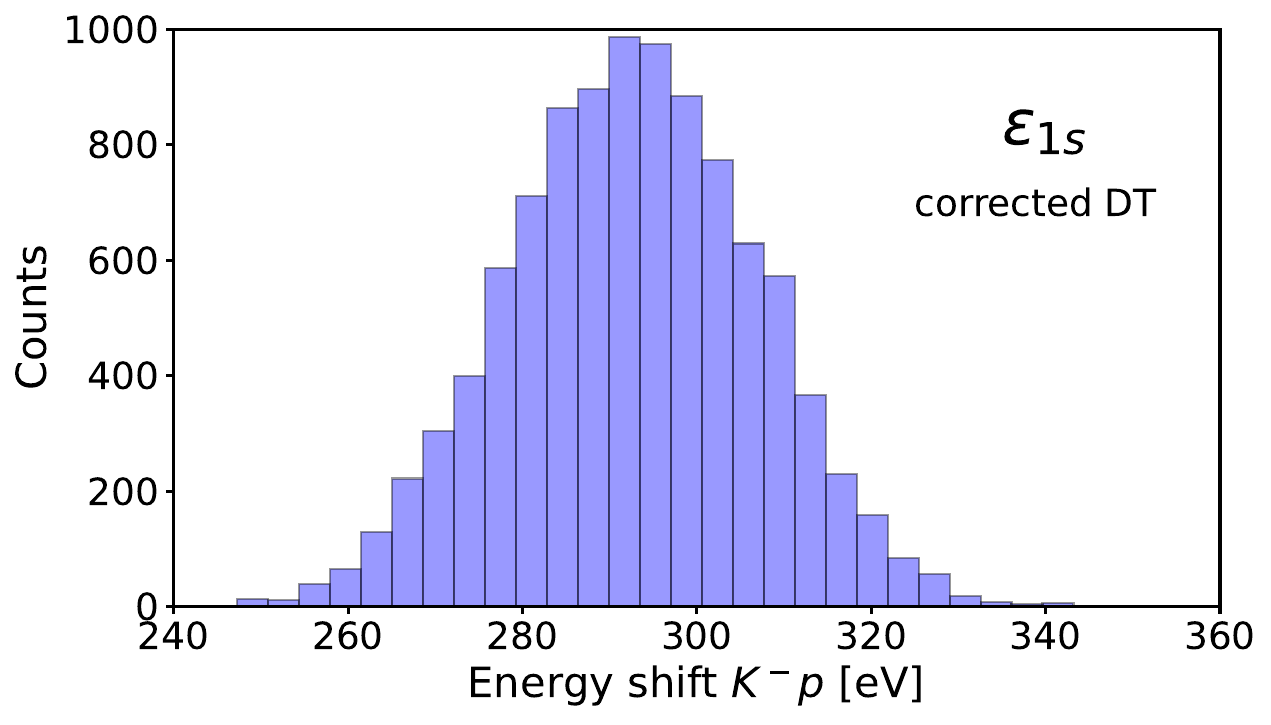}
    \hspace{0.005\textwidth}
    \includegraphics[width=0.39\textwidth]{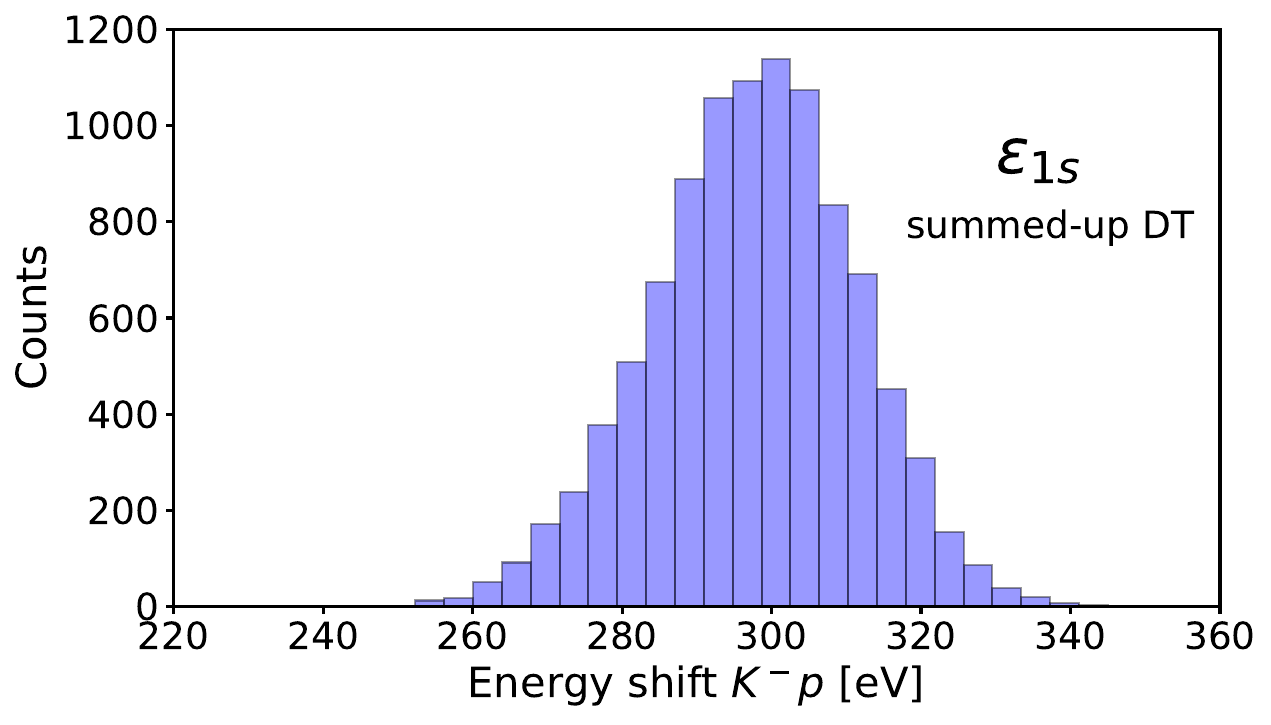}

    \vspace{-0.0em}

    \includegraphics[width=0.39\textwidth]{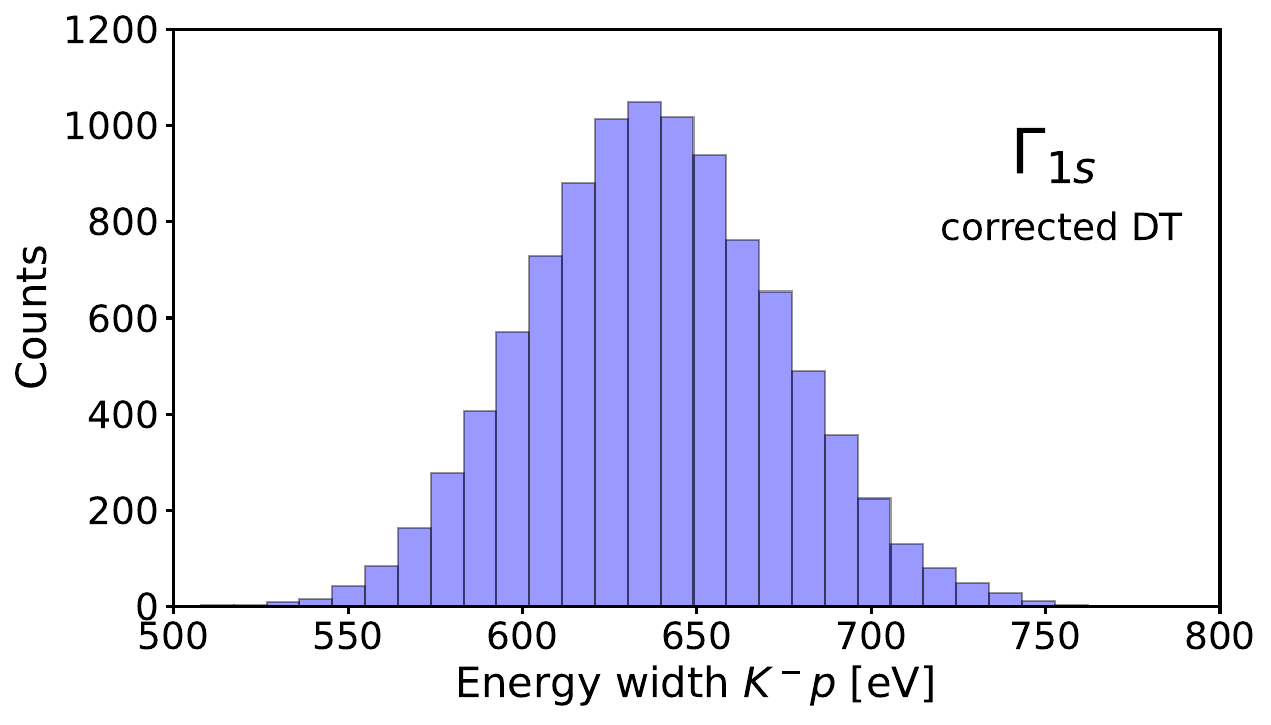}
    \hspace{0.005\textwidth}
    \includegraphics[width=0.39\textwidth]{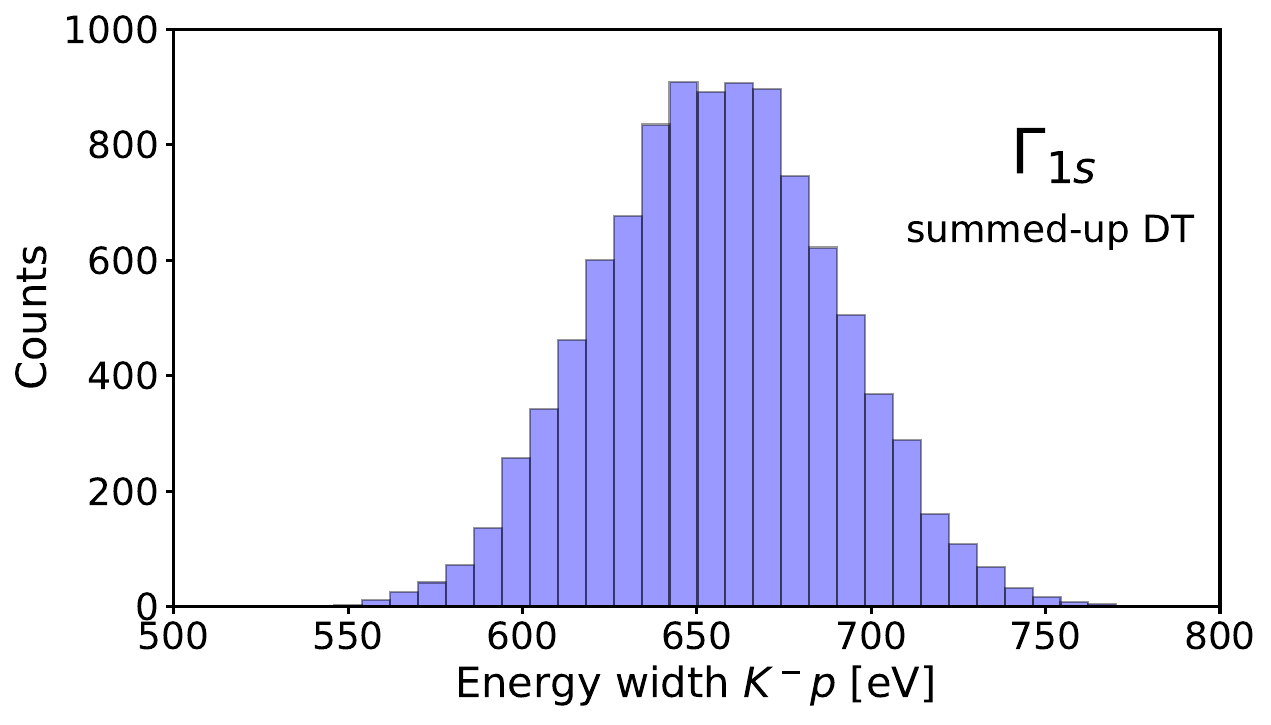}

    \vspace{-0.0em}

    \includegraphics[width=0.39\textwidth]{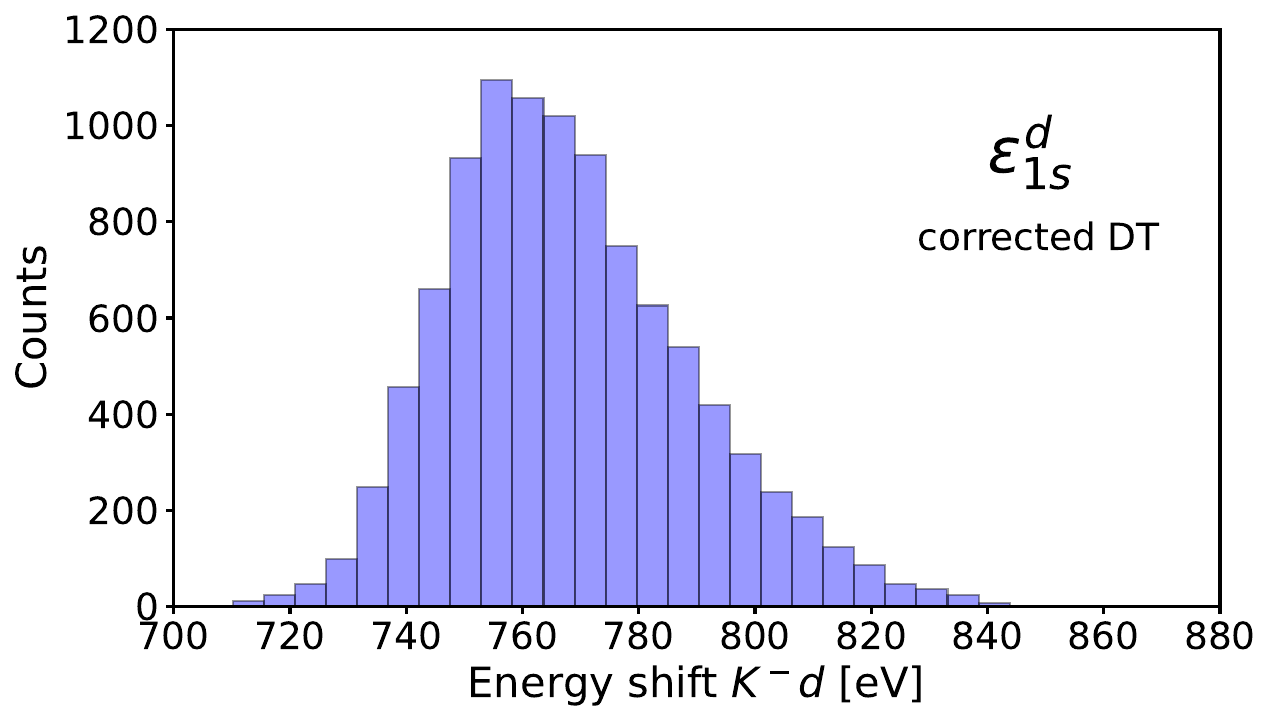}
    \hspace{0.005\textwidth}
    \includegraphics[width=0.39\textwidth]{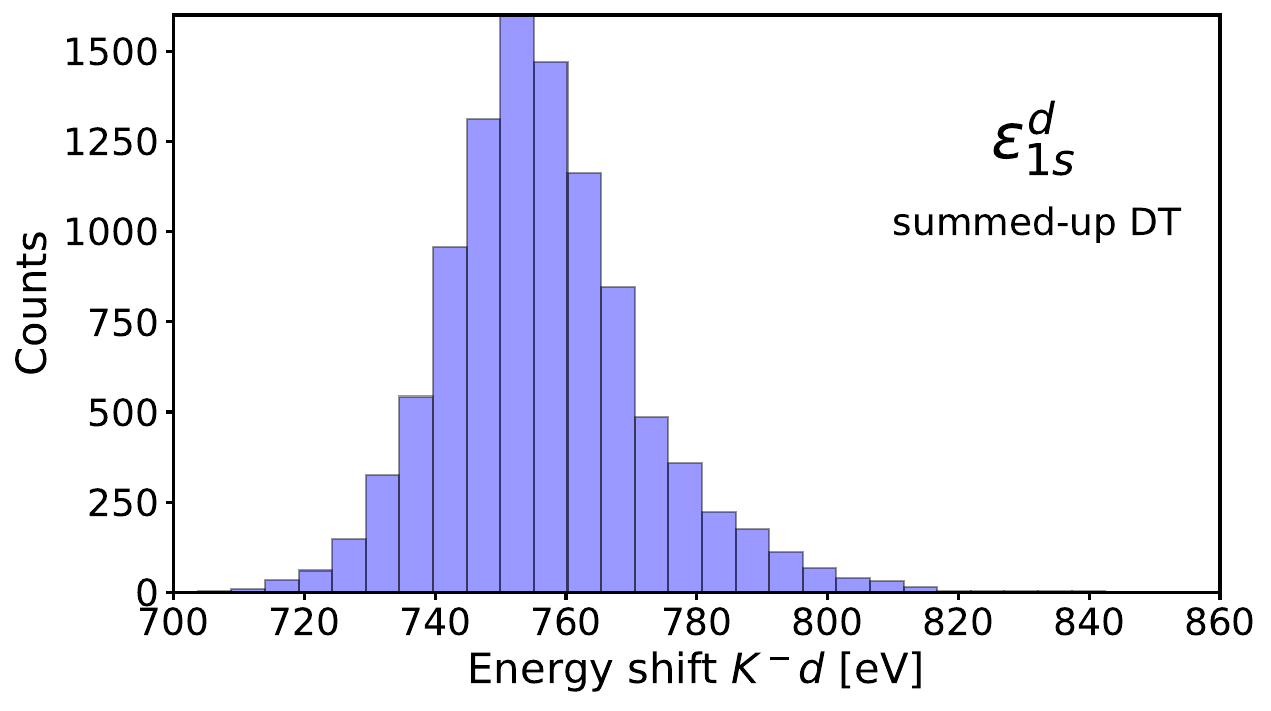}

    \vspace{-0.0em}

    \includegraphics[width=0.39\textwidth]{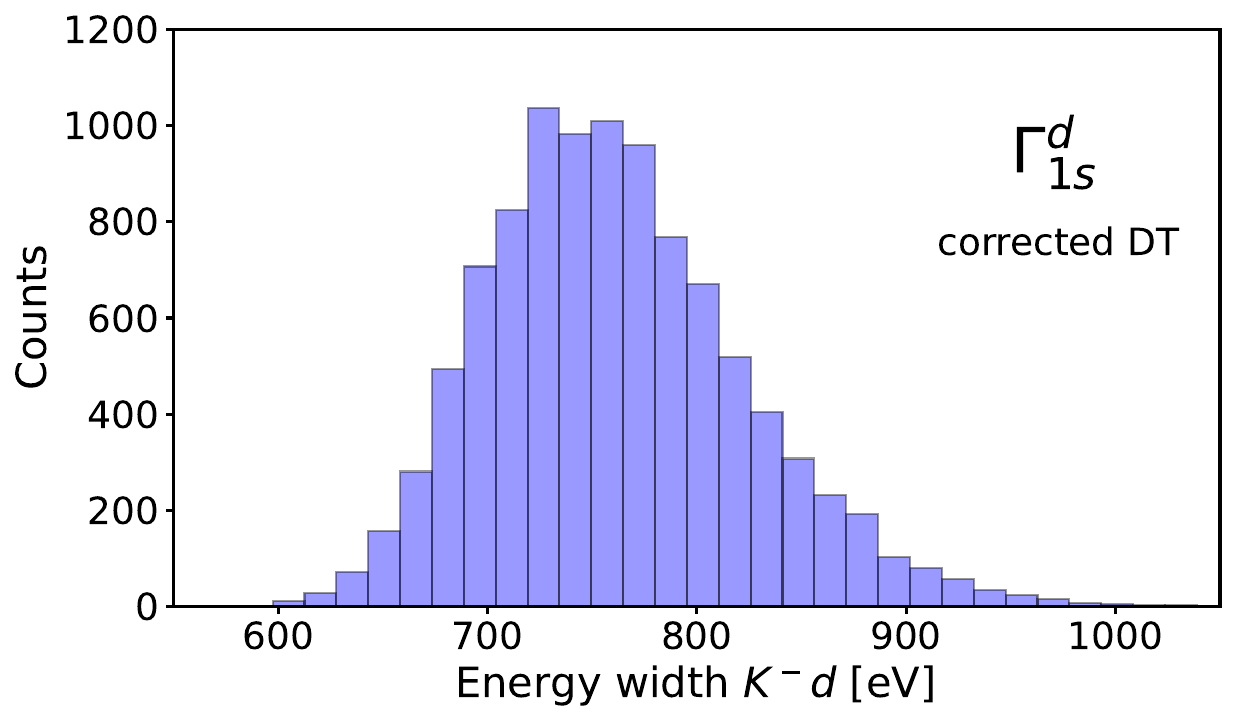}
    \hspace{0.005\textwidth}
    \includegraphics[width=0.39\textwidth]{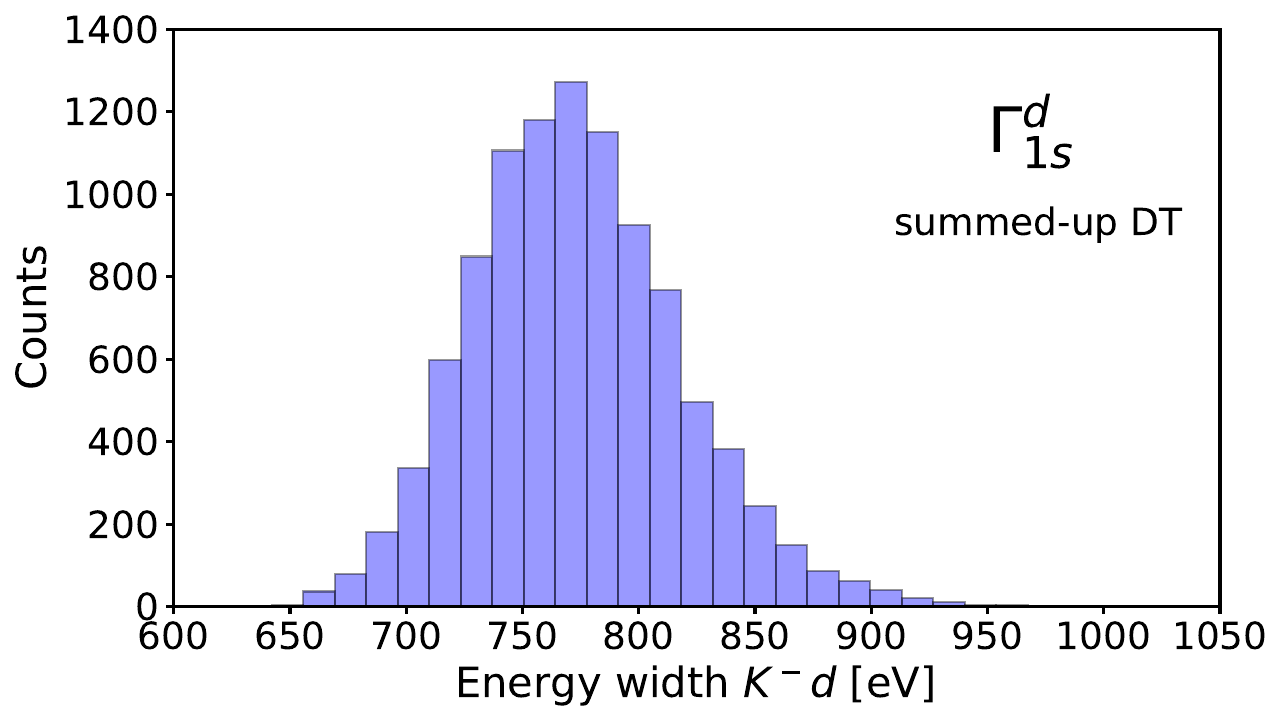}

\caption{
Distributions of the shifts and widths of kaonic hydrogen (first two rows) and kaonic deuterium (last two rows) obtained from the 10,000 fitted parameterizations, considering the corrected (left panels) or summed-up (right panels) Deser-Trueman expression.} \label{fig:shift_width_distributions}
\end{figure*}

Before proceeding with the parameter analysis, it is useful to examine the distributions of the shifts and widths obtained from the 10,000 fitted parameterizations. 
These are represented in 
Fig.~\ref{fig:shift_width_distributions} for the two considered Deser-Trueman prescriptions, using 30 bins for each histogram. As expected from the construction of the procedure, the observed shapes are consistent with the local quadratic behavior of the objective function around its minimum, which leads to approximately Gaussian variations of the fitted observables under small fluctuations of the input data.

\begin{figure*}[p]
    \centering

    \includegraphics[width=0.35\textwidth]{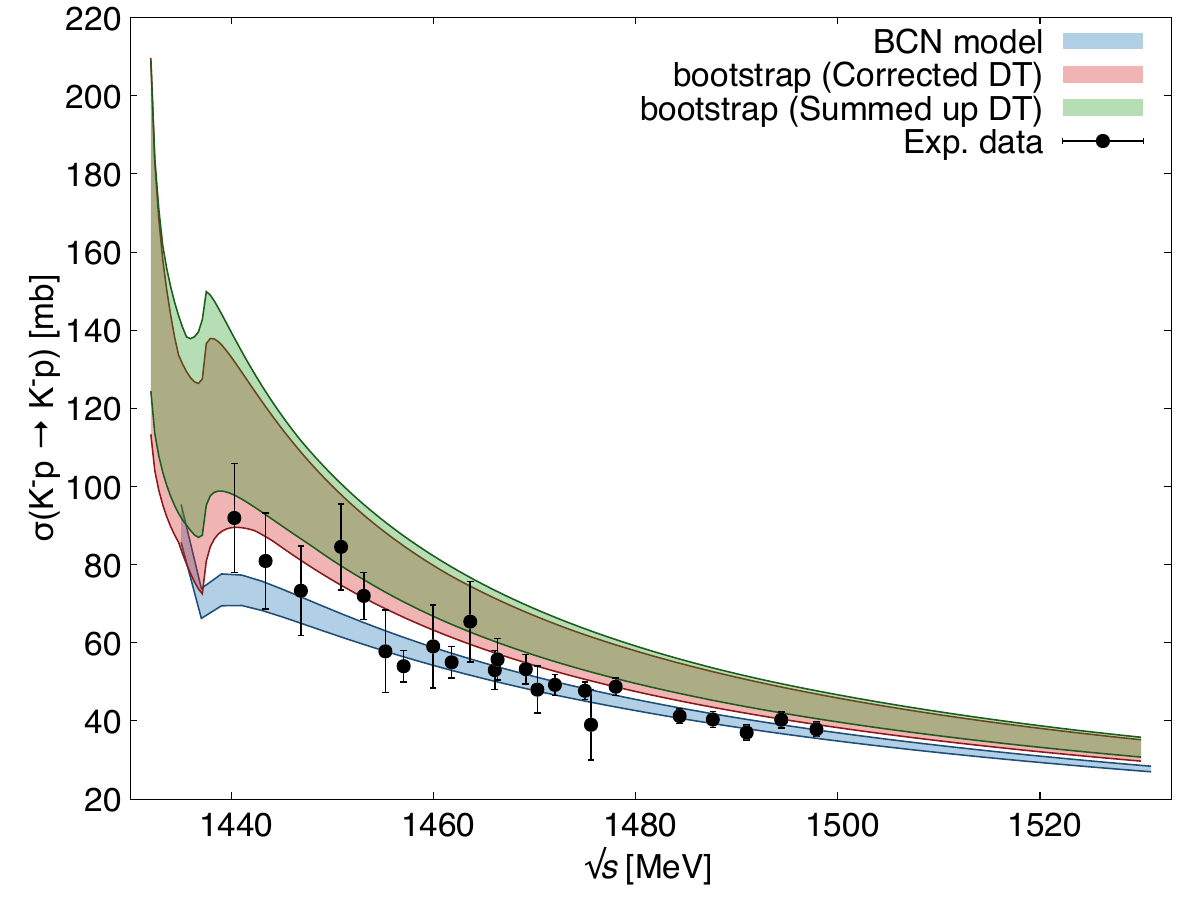}
    \hspace{0.005\textwidth}
    \includegraphics[width=0.35\textwidth]{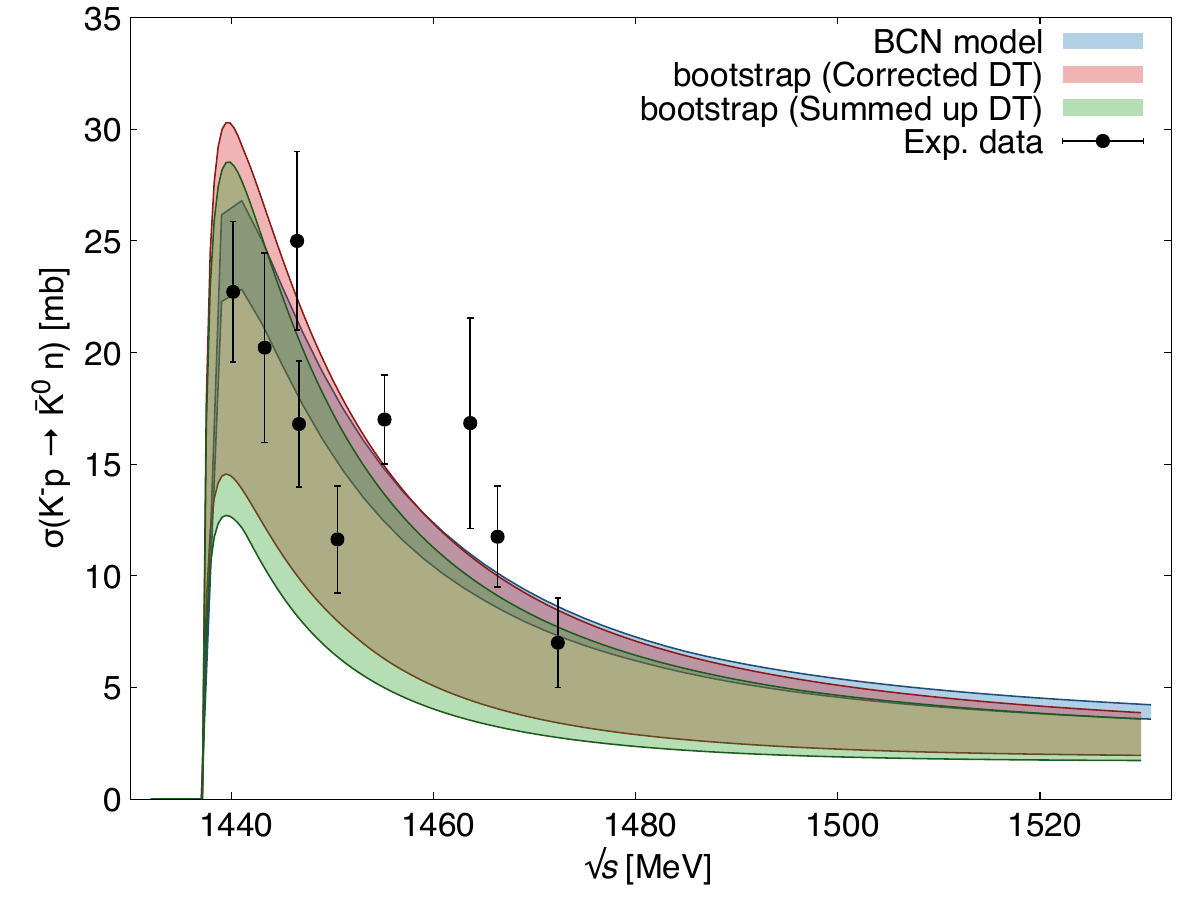}

    \vspace{-0.4em}

    \includegraphics[width=0.35\textwidth]{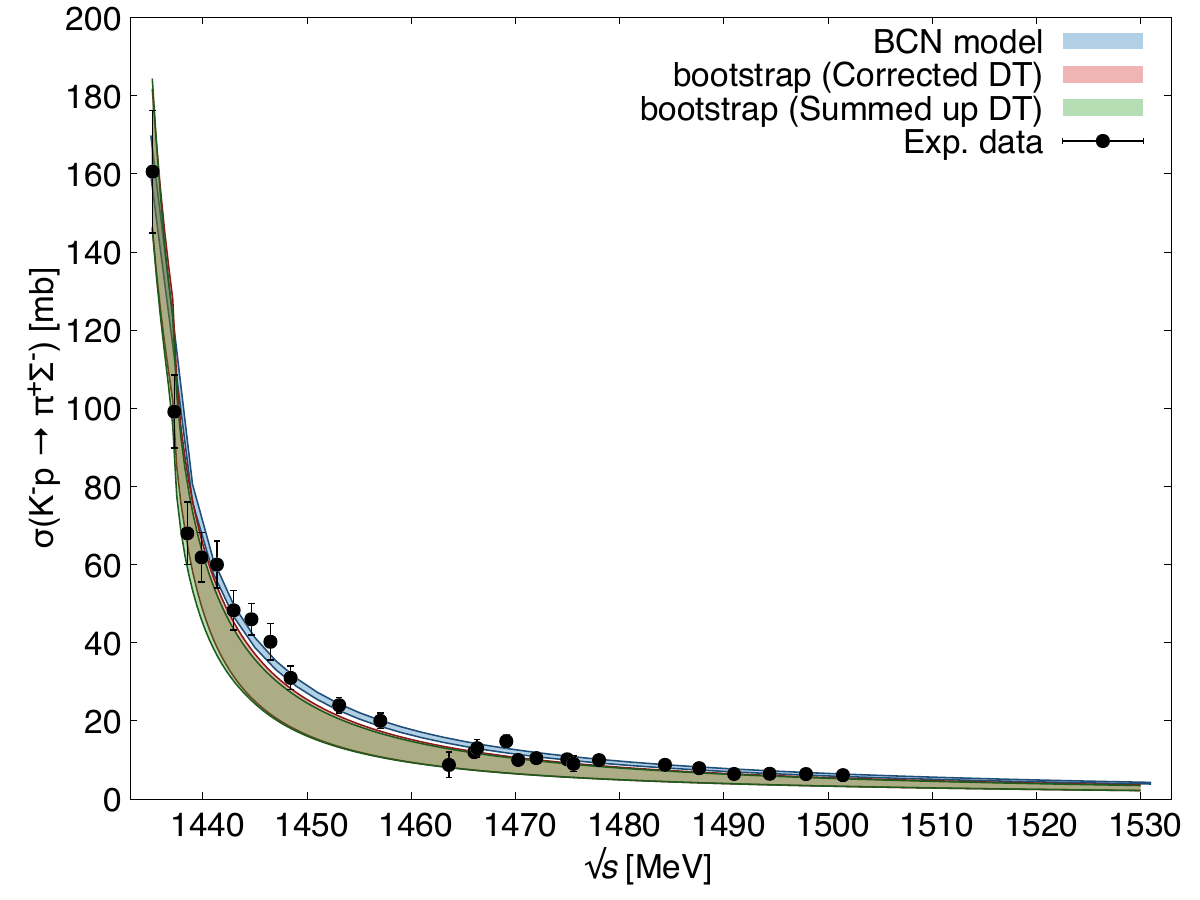}
    \hspace{0.005\textwidth}
    \includegraphics[width=0.35\textwidth]{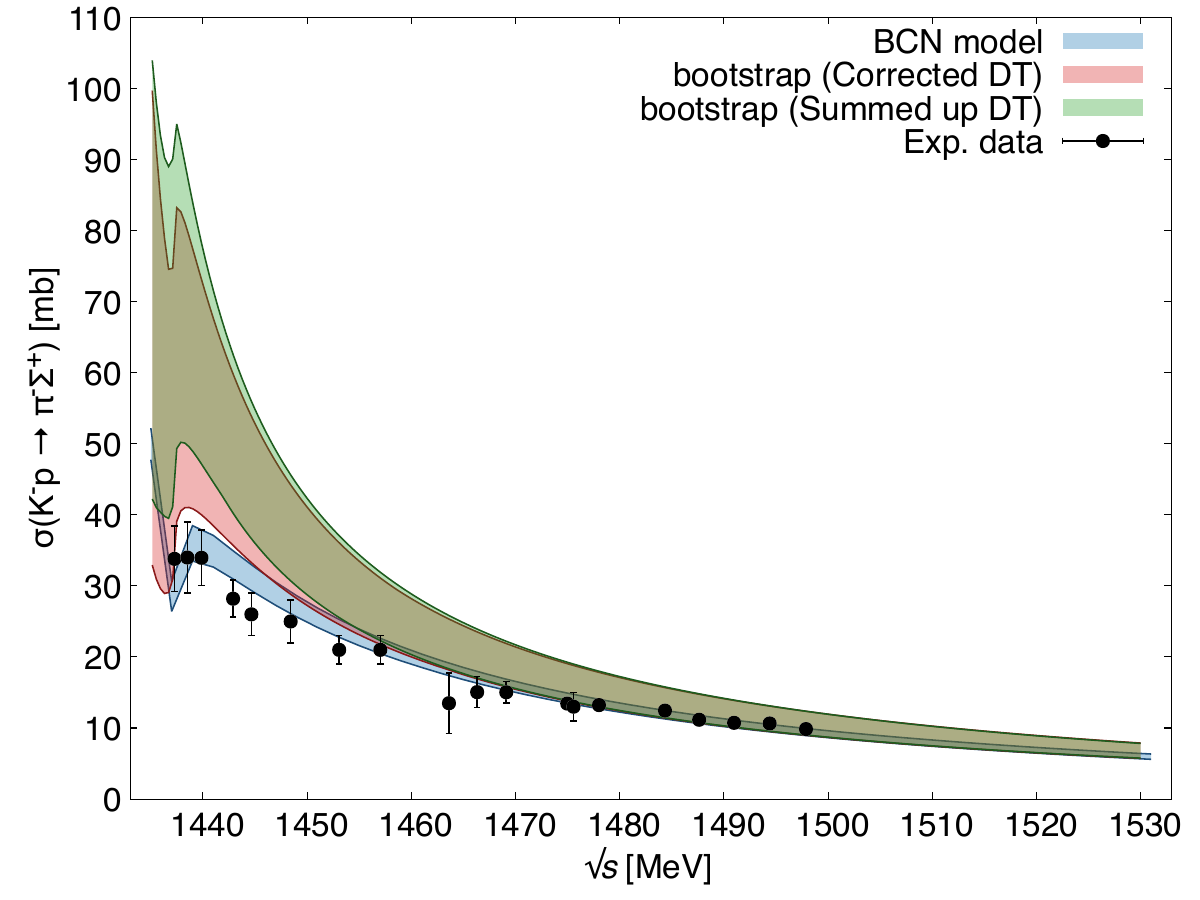}

    \vspace{-0.4em}

    \includegraphics[width=0.35\textwidth]{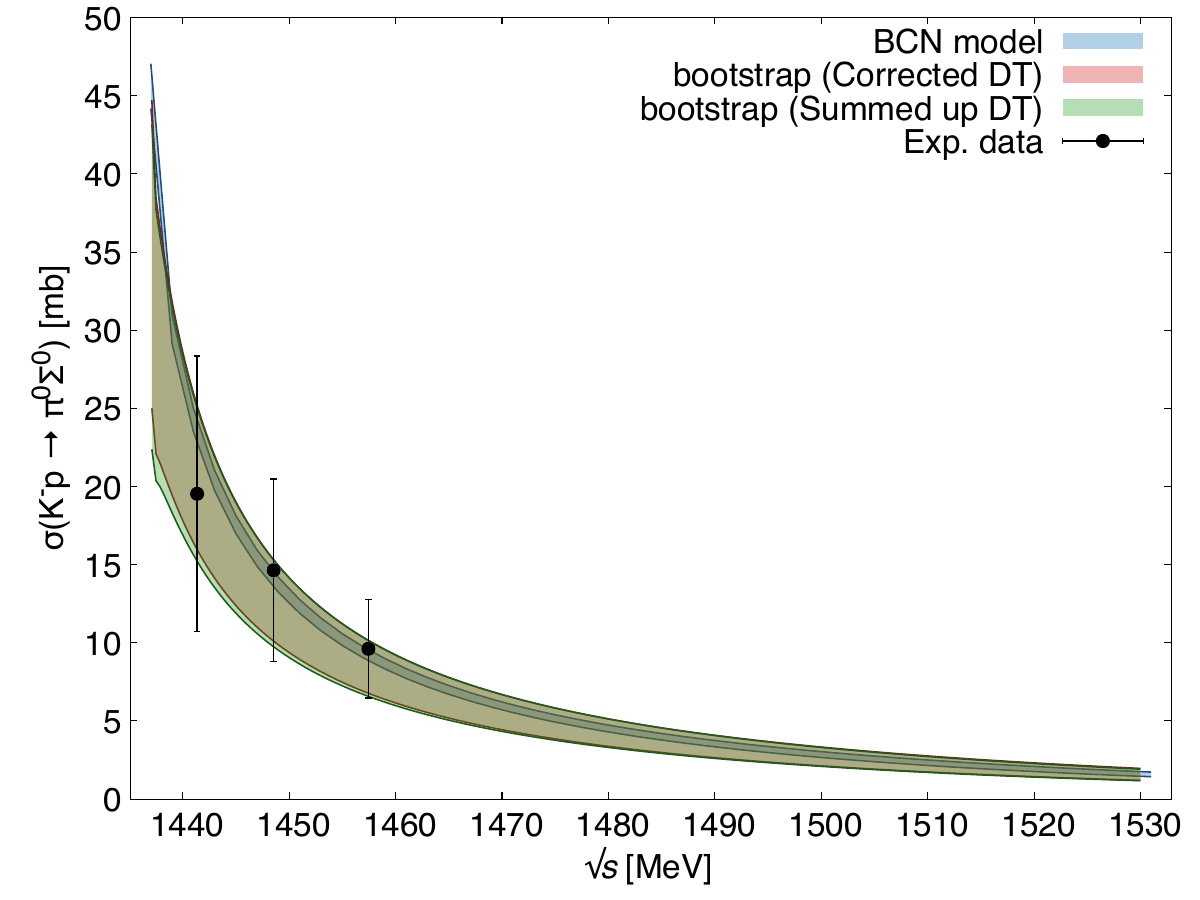}
    \hspace{0.005\textwidth}
    \includegraphics[width=0.35\textwidth]{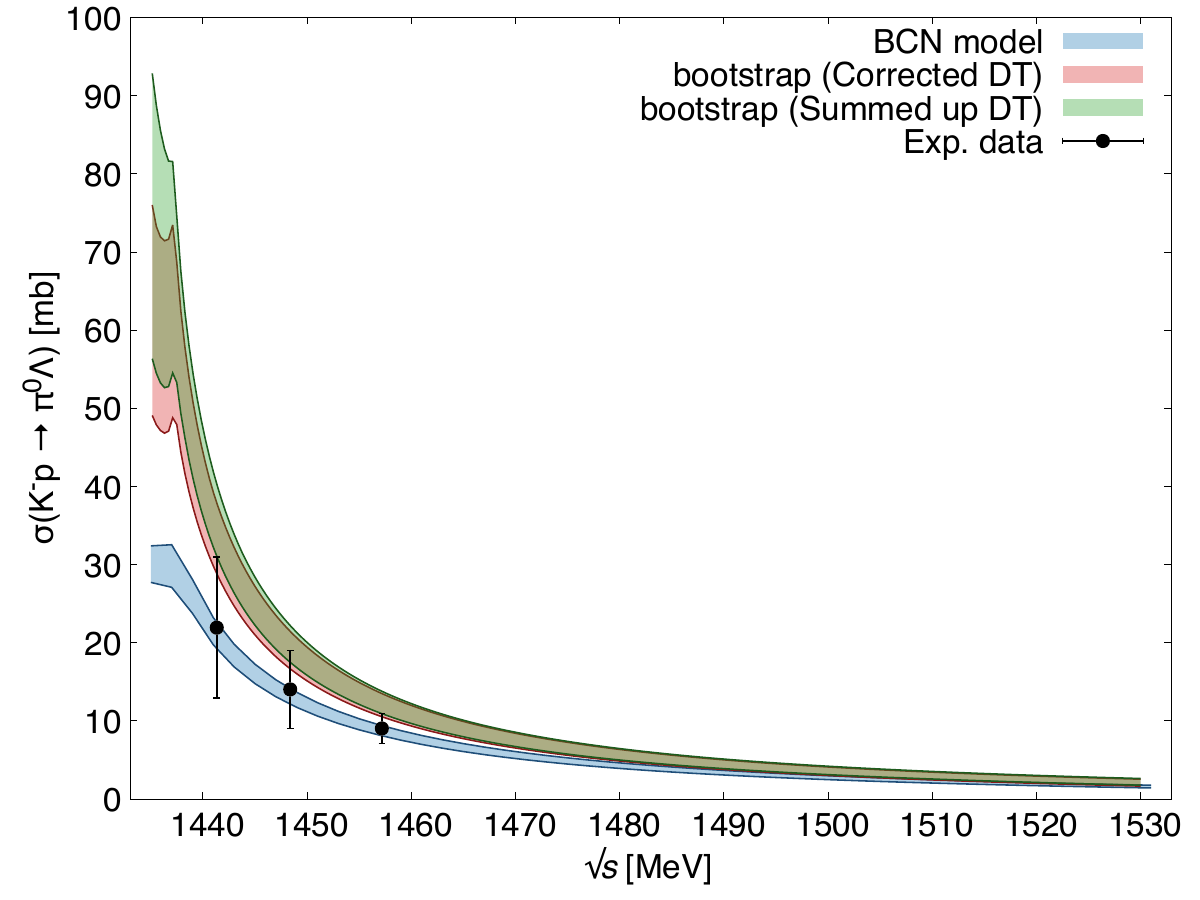}

    \vspace{-0.4em}

    \includegraphics[width=0.35\textwidth]{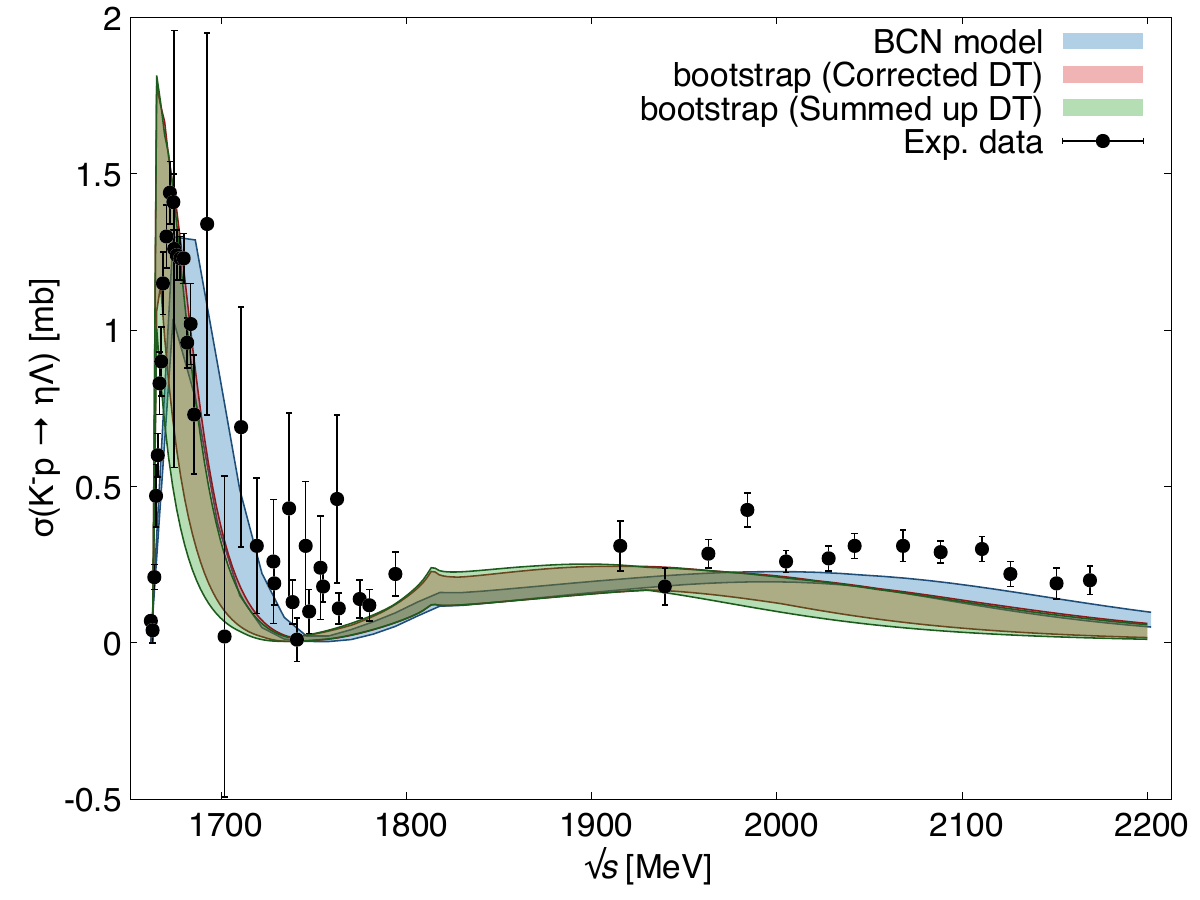}
    \hspace{0.005\textwidth}
    \includegraphics[width=0.35\textwidth]{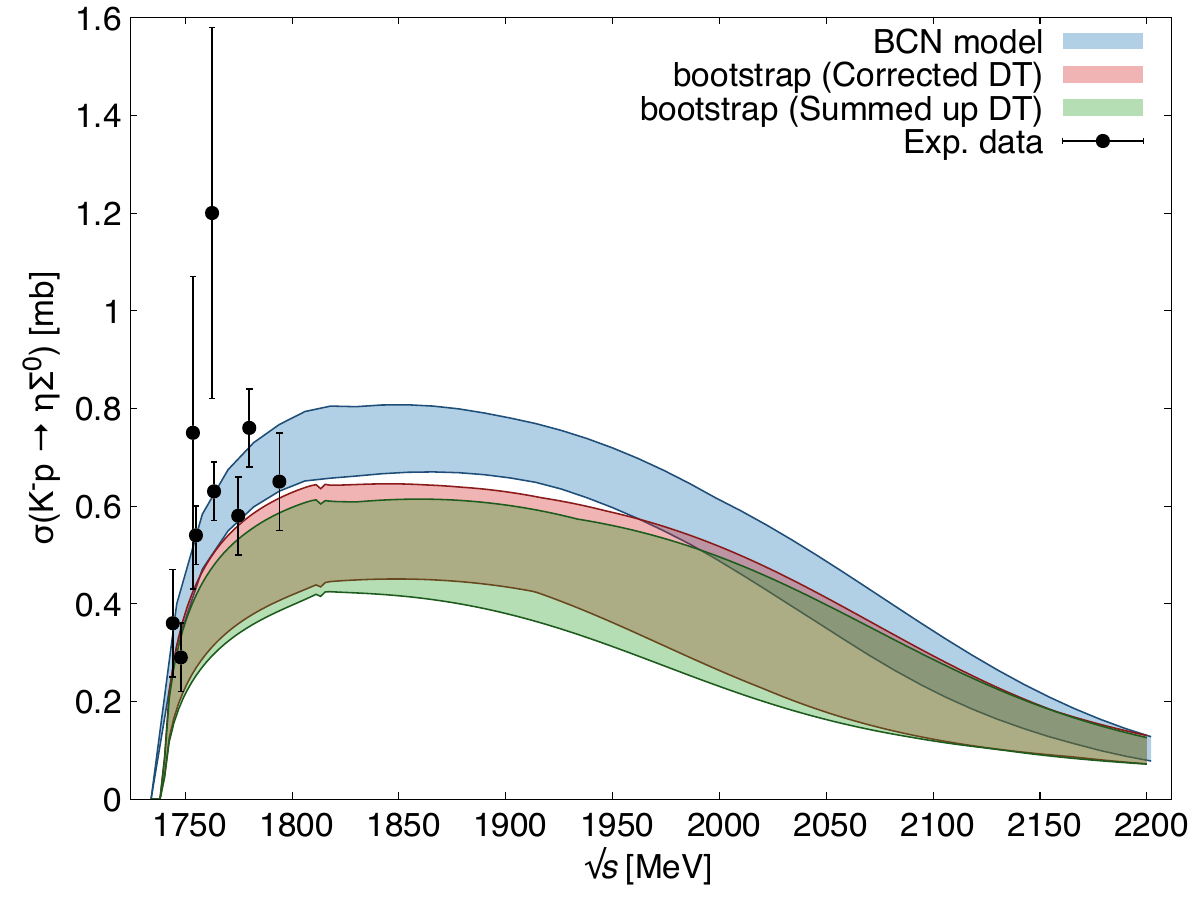}
    
        \vspace{-0.4em}

    \includegraphics[width=0.35\textwidth]{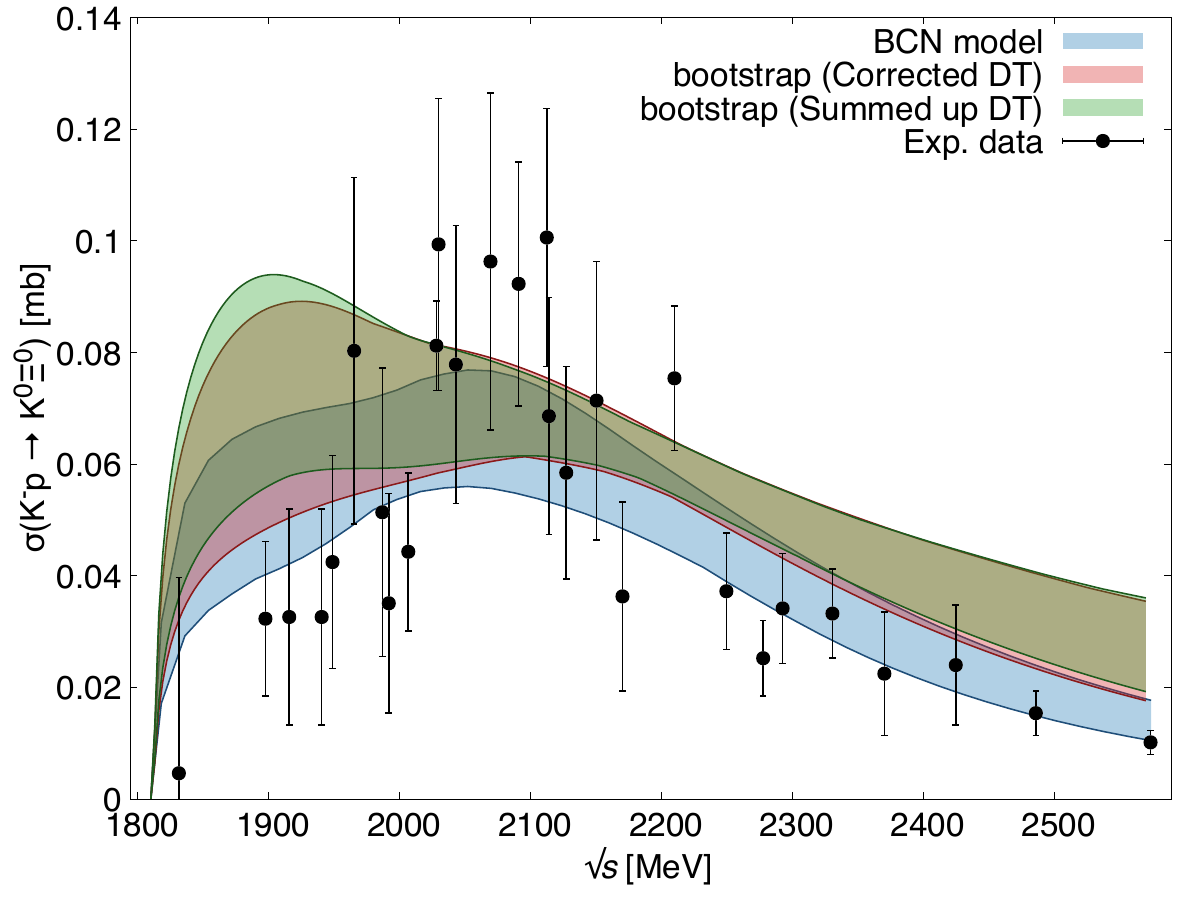}
    \hspace{0.005\textwidth}
    \includegraphics[width=0.35\textwidth]{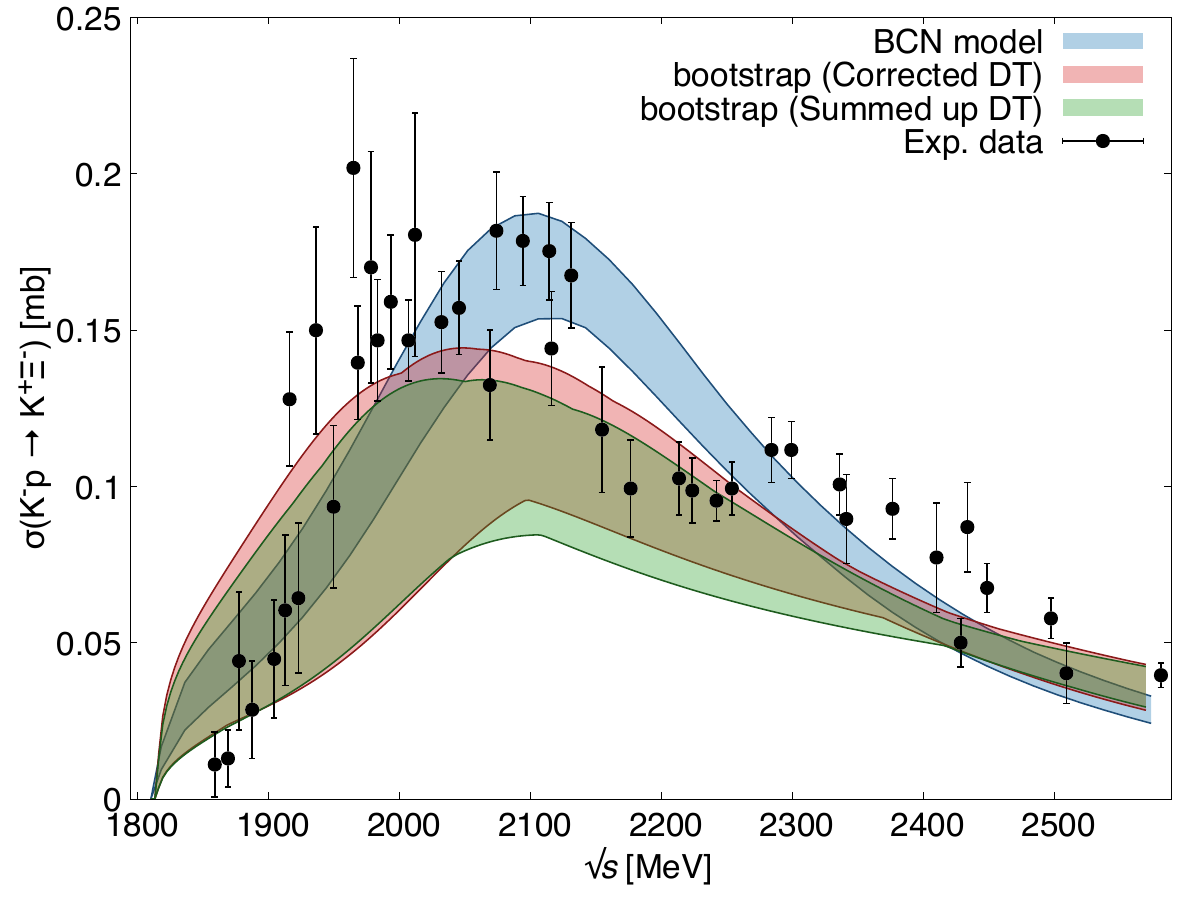}

\caption{Comparison among the error bands associated to the BCN model and both bootstrap analyses for the $K^-p \to K^-p$, $\bar K^0n$, $\pi^+ \Sigma^-$, $\pi^- \Sigma^+$, $\pi^0 \Sigma^0$, $\pi^0 \Lambda$, $\eta \Lambda$, $\eta \Sigma^0$, $K^+ \Xi^-$, $K^0 \Xi^0$ total cross sections. The experimental data are taken from Refs.~\cite{Kim:1965zzd,Birmingham-Glasgow-LondonIC-Oxford-Rutherford:1966onr,Trippe:1967wat,Trower:1968zz,London:1966zz,Burgun:1968ice,Dauber:1969hg,Dauber:1969hg,Carlson:1973td,Baxter:1973ggf,Jones:1974at,Berthon:1974kr,Mast:1975pv,Bangerter:1980px,Ciborowski:1982et,CrystalBall:2001uhc}.}

    \label{fig:xsections}
\end{figure*}

As for the analysis of the parameterizations, we first note that
the vast majority of the mean values of the parameters extracted from the bootstrap lie well within the original uncertainties of the BCN model (see Table~II of Ref.~\cite{Feijoo:2018den}), i.e., within the range $[p_i^{\rm BCN}-\delta p_i^{\rm BCN},p_i^{\rm BCN}+\delta p_i^{\rm BCN}]$, with the exception of two low-energy constants, namely $b_0$ and $d_4$. These two constants affect different contributions to the interaction: $b_0$ appears in one of the terms responsible for explicit chiral symmetry breaking, while $d_4$ enters the $p^2$-dependent NLO corrections. In particular, when performing the bootstrap using the corrected Deser-Trueman prescription, we obtain $b_0=0.076 \pm 0.009$ and $d_4=0.287 \pm 0.007$ $\rm GeV^{-1}$, while the corresponding values for the BCN model are $b_0=0.129\pm 0.032$ and $d_4=0.249^{+0.027}_{-0.033}$ $\rm GeV^{-1}$. 
This behavior is systematically reproduced by the summed-up Deser-Trueman bootstrap, for which we obtain $b_0=0.067 \pm 0.010$ and $d_4=(0.292 \pm 0.007)$ $\rm GeV^{-1}$.
The observed deviation should not be interpreted as evidence that $b_0$ and $d_4$ are, individually, particularly sensitive to the new SIDDHARTA-2 observables. Rather, it reflects the fact that the new experimental constraints induce a systematic displacement of the preferred parameter values relative to those of the original BCN fit. Within the present fitting prescription, this displacement is accommodated primarily through variations of $b_0$ and $d_4$, which therefore absorb a substantial part of the required adjustment. The fact that a similar shift is observed for both Deser-Trueman prescriptions further indicates that the effect is primarily driven by the new experimental constraints rather than by the particular Deser-Trueman prescription employed in the bootstrap. More generally, this behavior illustrates how the new constraints modify the preferred region of parameter space and the pattern of parameter variations through which the fit accommodates the new kaonic-atom observables.

The extent of these constraining effects is illustrated by the $K^-p \to K^-p$, $\bar K^0n$, $\pi^+ \Sigma^-$, $\pi^- \Sigma^+$, $\pi^0 \Sigma^0$, $\pi^0 \Lambda$, $\eta \Lambda$, $\eta \Sigma^0$, $K^+ \Xi^-$, $K^0 \Xi^0$ total cross sections, whose experimental data were included in the original BCN model fit. Figure~\ref{fig:xsections} compares, for each of these channels, the error bands obtained from the two bootstrap analyses with that of the BCN model. It can be appreciated that the bootstrap bands do not fully encompass the spread of the experimental data points, although they lie very close to them, with the exception of the $K^-p \to \pi^- \Sigma^+$ cross section at low energies.
Regarding their compatibility with the BCN bands, several energy regions exhibit no overlap with the original model predictions. This is particularly evident for transitions in which the interplay between the isospin components of the ${\bar K}N$ interaction is expected to be significant, as well as for channels driven by the isovector component. Representative examples of these two situations are provided by the $K^-p \to \eta \Sigma^0$ cross section, which is purely dominated by the $I=1$ component, and the $K^-p \to K^+ \Xi^-$ one, which displays a distinct isospin-dependent pattern within the BCN model. 
The bootstrap bands of the $K^-p \to \eta \Sigma^0$ cross section lie systematically below the BCN band, with only a small region of overlap at higher energies. This indicates that the parameter space spanned by the bootstrap parameterizations leads to a substantially different strength of the isovector component of the ${\bar K}N$ interaction. Furthermore, the right panel of Fig.~9 in Ref.~\cite{Feijoo:2018den}~\footnote{This figure shows the isospin-component distributions to the $K^-p \to K^+ \Xi^-$ cross section obtained from the BCN model (WT+Born+NLO in the figure legend). The $I=0$ distribution is concentrated at low energies, whereas the $I=1$ component peaks around $2180$~MeV.} can be compared with the last panel of our Fig.~\ref{fig:xsections}. This comparison reveals a clear reduction in the strength of the $I=1$ component for the bootstrap parameterizations, as the peak of the total cross section is shifted by almost $100$~MeV towards lower energies, where the $I=0$ component becomes dominant. Different isospin-dependent patterns from the original BCN model can also be expected for the bootstrap  $\bar K N \to\pi \Sigma$ cross sections, albeit to varying degrees. Together, these results indicate that the new SIDDHARTA-2 measurements provide additional constraints on the $\bar{K}N$ interaction and, in particular, increase the sensitivity to its poorly constrained isovector component. The simultaneous information from kaonic hydrogen and kaonic deuterium leads to a readjustment of the BCN parameter space that is reflected in several isospin-sensitive scattering observables. The effects observed in the present bootstrap analysis therefore illustrate the potential of the new kaonic-atom measurements to constrain the low-energy $\bar{K}N$ interaction beyond the information provided by kaonic-hydrogen data alone.

\section{Conclusions~\label{sec:conclusions}}

In this work, we have investigated the low-energy $K^-d$ interaction starting from chiral $\bar{K}N$ dynamics and extending the Fixed-Center Approximation to incorporate the coherent multiple-scattering contributions required by exact elastic unitary constraints. The resulting three-body framework is formulated in coupled channels and accounts for the charge-exchange mechanism, as well as isospin-breaking effects through the use of physical hadron masses. This formulation provides a consistent description of the low-energy $K^-d$ amplitude and allows the determination of both the $K^- d$ scattering length and the effective range.

Using the lowest-order OR and the next-to-leading-order BCN chiral models as input, we find that the unitary-restoring contributions reduce the predicted $K^-d$ scattering length by a factor of two with respect to the FCA result. We also obtain a small difference in the value of the scattering length and a moderate variation of the effective range when the charge-exchange processes are removed from the calculation. This indicates a non negligible role of these processes in kaon-deuteron scattering.

The restoration of unitarity also affects the obtained values of the kaonic deuterium energy shift and width. Using the corrected or summed-up Deser-Trueman expressions, the large values of these quantities in both OR and BCN models under the FCA get substantially reduced. In any case, the obtained energy shift in both models cannot describe the recent results from SIDDHARTA-2 experimental collaboration, while the width is compatible with the result of the OR model, and almost compatible with the BCN one. 
These results demonstrate the relevance of coherent multiple-scattering effects in the treatment of the kaon-deuteron system and provide a more consistent framework for confronting chiral $\bar{K}N$ interactions with kaonic-deuterium observables.

The resulting $K^-d$ amplitudes were also used to test the response of the BCN interaction to the recent SIDDHARTA-2 measurements of kaonic hydrogen and deuterium. A bootstrap analysis was performed by generating pseudo-data according to the experimental uncertainties and refitting the BCN parameterization, with parameter variations constrained by the uncertainties of the original fit, but without imposing correlations from the full covariance matrix. It is found that the simultaneous inclusion of hydrogen and deuterium observables modifies the BCN parameters and accommodates the new measurements within the flexibility of the parameterization. 

Both the corrected and summed-up Deser-Trueman prescriptions were considered for evaluating kaonic-atom observables. Their differences are relatively small and do not qualitatively alter the bootstrap conclusions, although they visibly affect the extracted observables. Since the original BCN parameterization employed the corrected Deser-Trueman prescription, this comparison estimates the sensitivity to the electromagnetic-correction treatment rather than demonstrating the superiority of either prescription.

The modifications induced by the new kaonic-atom data affect several $\bar{K}N$ scattering observables, particularly those sensitive to the isospin structure of the interaction. The combined hydrogen and deuterium information thus probes aspects of $\bar{K}N$ dynamics less constrained by kaonic-hydrogen data alone, providing additional insight into the isovector component. Overall, these results illustrate the constraining potential of the new kaonic-atom measurements and the importance of a consistent treatment of the $K^-d$ system in extracting low-energy antikaon-nucleon dynamics.

A definitive assessment of the impact of the new measurements requires a global fit in which the complete experimental data set---including the kaonic-atom measurements, subthreshold information from decay fractions, available scattering data, and the new femtoscopic measurements---is treated simultaneously, with the full parameter correlations consistently taken into account. Such an analysis would quantify in a statistically consistent manner to what extent the SIDDHARTA-2 measurements can reduce the remaining parameter and isospin ambiguities of the ${\bar K}N$ interaction.

\begin{acknowledgments}

This work has been supported by the project numbers CEX2019-000918-M (Unidad de Excelencia ``Mar\'ia de Maeztu'') and PID2023-147112NB-C21, financed by the Spanish MCIN/ AEI/10.13039/501100011033/; and by Contract 2021 SGR 171 by the Generalitat de Catalunya.
JMT-R also thanks Grant No. 402942/2024 by the Brazilian CNPq (National Council for Scientific and Technological
Development). This work is also part of the Grants {\small PID2023-147458NB-C21} and {\small CEX2023-001292-S} funded by {\small MICIU/AEI/10.13039/501100011033} and by ERDF/EU, as well as of the Grant {\small CIPROM/2023/59} funded by Generalitat Valenciana {\small 10.13039/501100003359}.
\end{acknowledgments}

\bibliography{references}

\end{document}